\documentclass[12pt, twocolumn,twocolappendix]{aastex631}

\usepackage[english]{babel}
\usepackage{amsmath,amssymb,amsfonts, bm,bbm,slashed, subdepth}
\usepackage{mathalfa}
\usepackage{graphicx}
\usepackage{hyperref}
\usepackage{enumerate}
\usepackage{epsfig, subfigure}
\usepackage{multirow}
\usepackage{booktabs}

\usepackage{braket}

\allowdisplaybreaks

\DeclareMathAlphabet\mathbfcal{OMS}{cmsy}{b}{n}

\let\oldhat\hat

\renewcommand{\hat}[1]{\bm\oldhat{\mathbf{#1}}}

\newcommand{\lss}[1]{\textcolor{black}{#1}}

\newcommand{\gff}{g_\text{ff}^{(Q)}}
\newcommand{\gid}{g_\text{id}}
\newcommand{\gnonid}{g_\text{non-id}}
\newcommand{\gint}{g_\text{ex}}

\newcommand{\ie}{\textit{i.e.}}
\newcommand{\eg}{\textit{e.g.}}

\newcommand{\tablesepvert}{0.83}

\newcommand{\TableComment}{published in its entirety in the machine-readable format with increased range, precision, and finer spacing.}

\def\Plus{$+$}
\def\Minus{$-$}
	
\begin{document}

\title{Accurate Gaunt factors for non-relativistic quadrupole bremsstrahlung}

\author[0000-0002-9165-4813]{Josef Pradler}
\affiliation{Institute of High Energy Physics, Austrian Academy of Sciences, Nikolsdorfergasse 18, 1050 Vienna, Austria}
\email{josef.pradler@oeaw.ac.at}
\author[0000-0002-7059-2094]{Lukas Semmelrock}
\affiliation{Institute of High Energy Physics, Austrian Academy of Sciences, Nikolsdorfergasse 18, 1050 Vienna, Austria}
\email{lukas.semmelrock@oeaw.ac.at}

\begin{abstract}
    The exact result for non-relativistic quadrupole bremsstrahlung in a Coulomb field was established only recently in~\cite{Pradler:2020znn}. It requires the evaluation and integration of hypergeometric functions across a wide range of parameters and arguments, which, in practice, is unfeasible. Here we provide a highly accurate tabulation of the Gaunt factor for quadrupole radiation,  its thermal average in a Maxwellian plasma, and the associated cooling function over the entire kinematically relevant range. In addition, we provide a simple approximate formula for the emission cross section which works to within a few percent accuracy for all practical purposes.
    The results can be applied to the scattering of electrons with themselves, for which quadrupole radiation is the dominant process. 
\end{abstract}

\section{Introduction}

The elementary process of bremsstrahlung is ubiquitous in nature and of ample relevance in many branches of physics. The leading order process in the relative non-relativistic velocity of two colliding bodies of differing charge-to-mass ratio is dipole emission. Its classical description is due to~\cite{Kramers:1923} and it is used in the definition of a Gaunt factor, \ie, the factor that takes the limiting classical expression to the correct quantum mechanical one~\citep{1930RSPTA.229..163G}. The \lss{exact non-relativistic quantum mechanical result was obtained by~\cite{Sommerfeld1931}} and as such, it contains both, the Born and classical limits; the Gaunt factor hence approaches  unity in the latter limit. The free-free dipole emission process is of course of ample importance in astrophysics and a large body of literature exists on it; treatments of non-relativistic electron-ion bremsstrahlung include \citep{1935MNRAS..96...77M, 1961ApJS....6..167K, 1972ApJ...174..227J, 1975ApJ...199..299K, 1988ApJ...327..477H, vanHoof:2014bha, Weinberg:2019mai,Chluba:2019ser} and tabulations of the Gaunt factor enter, \eg, spectral synthesis codes such as~\texttt{CLOUDY}~\cite{2013RMxAA..49..137F,2017RMxAA..53..385F}.

In stark contrast and perhaps surprisingly,  the next order process in relative velocity, quadrupole brems\-strahlung, has received much less attention in the literature. This may partly be attributed to the considerably increased complexity of the calculation. For electrons, the Born result was obtained by~\cite{Fedyushin:1952hg} and extended to arbitrary spin by~\cite{1981PhRvA..23.2851G}.
\cite{Elwert:1939km} found a way to extend the validity of the Born results into the mildly classical regime by multiplying the cross sections by a ratio of Sommerfeld factors. This prescription was applied to electron-electron bremsstrahlung by~\cite{Maxon1967}; \lss{ 
further dedicated Born-level works  include~\cite{1975ZNatA..30.1099H,1975ZNatA..30.1546H,1989A&A...218..330H} and \cite{2002NCimB.117..359I}.} It is important to note, however, that the Elwert approximation fails eventually, and the full result must then be used.

The exact expression for the emission cross section was only obtained most recently by us in~\cite{Pradler:2020znn}. The central result can be framed in terms of a quadrupole tensor that gets evaluated in the mutual Coulomb field of the colliding non-relativistic particle pair. The final expression for the differential cross section is a lengthy linear combination of hypergeometric functions of the sort
 $   {}_2F_1(i\nu_f, i\nu_i;1;z)$
and their derivatives with respect to its argument~$z$. Here, 
\begin{align*} \label{eqn:nuif}
    \nu_{i,f} \equiv \frac{-Z_1Z_2\alpha}{v_{i,f}} \to 
    \begin{cases}
    \text{Born regime} & (|\nu_{i,f}|  \ll 1) , \\
    \text{classical regime} & (|\nu_{i,f} |  \gg 1 ) ,
    \end{cases}
\end{align*}
are the Sommerfeld parameters where
$\nu_{i,f}$ is positive (negative) for attractive (repulsive) interactions; $v_{i(f)}$ is the initial (final) relative velocity of the colliding particle pair with individual charges $Z_1$ and $Z_2$.
Despite the very complex analytical nature of the end result, we show that the ensuing cross sections reduce exactly to the previously known Born and classical limits. This attests to the correctness of the calculation.

In practical applications, however, it is not feasible to evaluate the integrand---which is a sum of many terms of products of hypergeometric functions---with sufficient accuracy in a reasonable amount of time. The calculation of hypergeometric functions of large imaginary arguments is a difficult task~\citep{2008CoPhC.178..535M,2016arXiv160606977J} and in the strict classical limit $\hbar\to 0$ the arguments of  $   {}_2F_1(i\nu_f, i\nu_i;1;z)$ all approach  $|\nu_{i,f}|\to \infty$ and $|z|\to \infty$.
It is the purpose of this paper, to provide an accurate tabulation of the  Gaunt factor across the entire kinematically relevant regime to facilitate usage of our results and to make them broadly available for applications.

The paper is organized as follows: 
in Sec.~\ref{sec:ff_gaunt} we introduce the Gaunt factor for quadrupole radiation and evaluate it to high precision over a large range in parameters for repulsive interactions.
In Sec.~\ref{sec:therm_gaunt} we provide thermal averages in a Maxwellian plasma, obtain the production spectrum and opacity.
Section~\ref{sec:tot_gaunt} deals with the cooling function. Conclusions are offered in~\ref{sec:conclusions}. The Appendix provides further details on our calculations.

Unless stated otherwise, we use natural units $\hbar = c = k_B =1$. In the occasional conversion to ordinary units, we use the CODATA~2018 recommended values of fundamental constants~\citep{codata}.

\section{Free-free Gaunt Factor}
\label{sec:ff_gaunt}

\begin{figure*}[tb]
		\includegraphics[width=\textwidth]{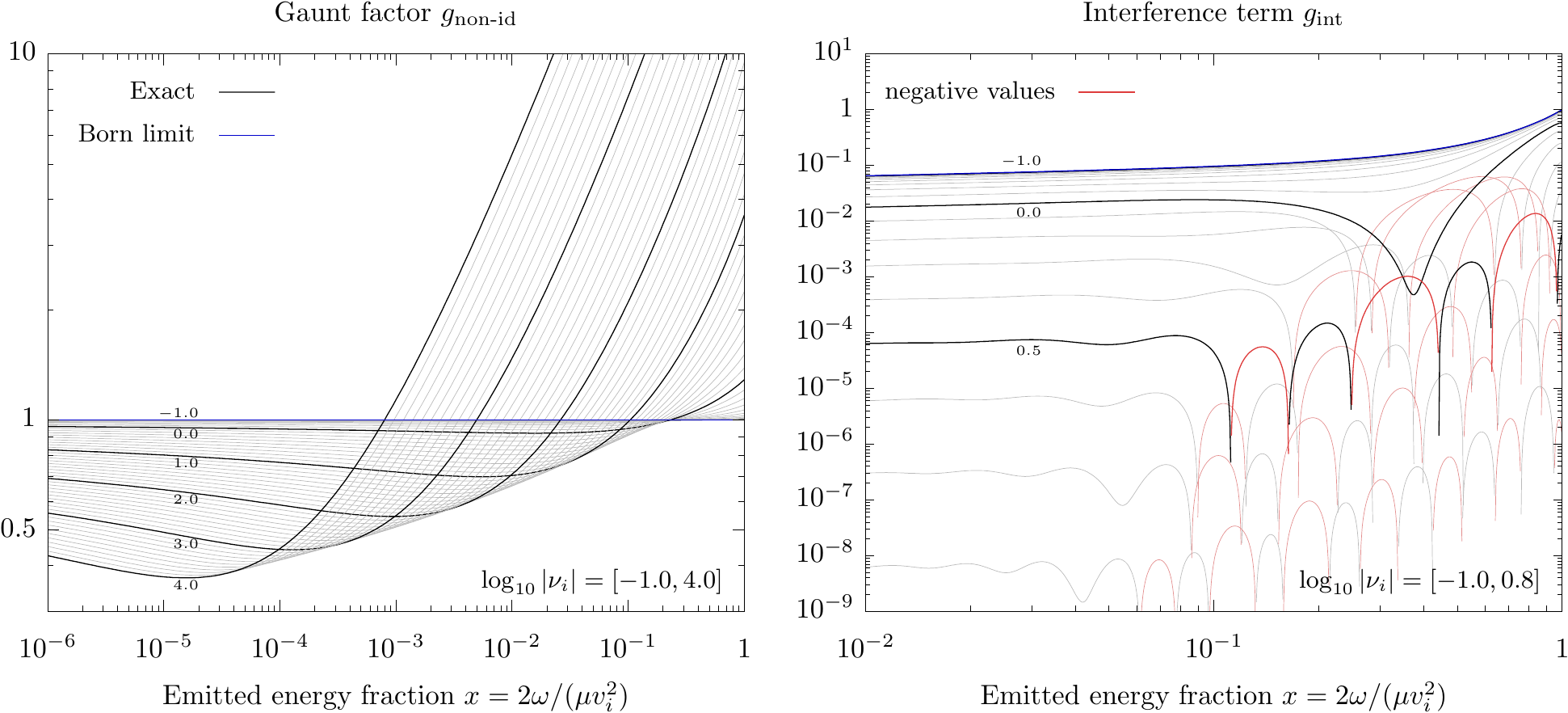}
		\caption{Free-free Gaunt factor for non-identical particles (left), and \lss{exchange} term (right) over a wide range of parameters $\nu_i$ and $x$; $\gnonid$ is invariant under a sign change of $\nu_i\to -\nu_i$ and hence applies for repulsive and attractive interactions alike; exchange effects captured in $\gint$ are not present for distinguishable particles $(Z_1\neq Z_2)$. The lines are shown in increments of $\log_{10}|\nu_i|=0.1$ with thicker lines as labeled. In the left panel, red line-sections correspond to negative values; Born limits are shown in blue. \label{fig:g_ff}}
\end{figure*}

\subsection{Definition of the Gaunt factor}
\label{sec:def_gaunt}

In this section we provide the definition of a Gaunt factor for the free-free quadrupole transition, $\gff$, which will be tabulated in the subsequent sections. We break with tradition and define $\gff$ as the multiplicative factor of the Elwert cross section and not of the classical expression as it is done for dipole emisson.%
\footnote{For the dipole case see~\cite{1961ApJS....6..167K,vanHoof:2014bha,Weinberg:2019mai,Chluba:2019ser}.
}
Hence, for the quadrupole case we define~\citep{Pradler:2020znn}
\begin{equation}
 \label{eq:GauntQuadrupole}
   \frac{d\sigma_{Q}}{dx}  = 
  	 e^{-2\pi \nu_i}\frac{S_f}{S_i}  \left.\frac{d\sigma_Q}{dx} \right|_{\substack{\text{Born}\\\text{non-id}}}
  	 \times \gff(x, \nu_i) ,
\end{equation}
where the left hand side is the exact differential quadrupole cross section for emitting a photon of fractional center-of-mass energy $x=2\omega/(\mu v_i^2)$ and the right hand side is the Elwert cross section for nonidentical particles times the Gaunt factor; the Sommerfeld factors are $S_{i,f} = \pm 2\pi \nu_{i,f} /(e^{ \pm 2\pi\nu_{i,f}}-1 )$ where the positive and negative signs correspond to~$i$ and $f$, respectively.%

The reasons for the definition in~\eqref{eq:GauntQuadrupole} are three-fold. First, for the important process of electron-electron bremsstrahlung, Coulomb corrections remain much smaller than for electron-ion bremsstrahlung, and the Born limit $|\nu_i|\ll 1$ is the more likely applicable case. Second, the Elwert correction factor extends the validity of the Born result to larger $|\nu_i|$ and covers hard photon ($x\to 1$) emission, since $|\nu_f| = |\nu_i|/\sqrt{1-x} $ becomes large in this limit. It then follows that the Gaunt factor remains a number close to unity across a maximal range that is of ample practical interest. 
Third, the Gaunt factor defined in this way has the symmetry property $\nu_i \to - \nu_i$ (to be further detailed below) and it allows one to treat attractive and repulsive interactions on the same footing.
\lss{In the numerical evaluation of Eq.~\eqref{eq:GauntQuadrupole}  and without deduction on accuracy one may replace the Elwert factor $\exp(-2\pi \nu_i) S_f/S_i$ in the following regions of parameter space by
\begin{align*}
%e^{-2\pi \nu_i}\frac{S_f}{S_i} \simeq 
\frac{1}{\sqrt{1-x}} \times
    \begin{cases}
    1 & \text{attractive}, \nu_i  > 10\\
    e^{-2\pi \nu_i y} & \text{repulsive, }  |\nu_i|>10 \land |\nu_i y | < 10, \\
    0 & \text{repulsive, } |\nu_i|>10 \land |\nu_i y | > 10.
    \end{cases} 
\end{align*}
with $y =1-1/{\sqrt{1-x}}$.
This avoids numerical under- or overflow in the exponentials when using double digit precision; in the repulsive case and for $|\nu_i y | > 10$ the factor is always smaller than $10^{-20}$ rendering the cross section negligible. 
}

 The non-relativistic Born cross section is determined by the symmetry properties of the colliding particles.%
 \footnote{In the Born limit, the difference between attractive and repulsive potentials is not resolved; spin is a relativistic concept and does not enter in leading order of relative velocity.}
For non-identical particles, such as in electron-ion scattering, the cross section reads~\citep{1990ApJ...362..284G}
	\begin{align} \label{eq:bornnonid}
		x & \left.\frac{d\sigma_Q}{dx} \right|_{\substack{\text{Born}\\\text{non-id}}}
		 =
		\frac{8\alpha^3 Z_1^2 Z_2^2 \mu^2}{15}
		\left(\frac{Z_1}{m_1^2} + \frac{Z_2}{m_2^2}\right)^2\nonumber\\ & \qquad \times 
		\bigg[10 \sqrt{1-x} 
		+ 3 (2-x)\ln \left(\frac{1+ \sqrt{1-x}}{1- \sqrt{1-x}}\right) \bigg] ,
	\end{align}
where $m_{1,2}$ are the masses of the colliding particles. In the scattering of identical particles, such as for electron-electron collisions, additional \lss{exchange} effects appear. From the definition~\eqref{eq:GauntQuadrupole} in terms of the non-identical Born cross section~\eqref{eq:bornnonid} it then follows that the Gaunt factor for identical particle scattering carries the additional contribution from the \lss{exchange}  terms. 
We write this~as
\begin{align}
\label{eq:gaunt_id}
    \gid  = 
     \gnonid
      + \frac{(-1)^{2s}}{2s+1} \gint .
\end{align}
For non-identical particle scattering $\gint=0$ and $\gff= \gnonid$, whereas for identical particle scattering $\gff= \gid$ with $s = 0,\ 1/2$ being the spin of the colliding identical particle pair.

Quadrupole bremsstrahlung is the leading  emission process for the scattering of identical particles for which the interaction is repulsive, $Z_1 Z_2 > 0$.
For attractive interactions, $Z_1 Z_2 < 0$, quadrupole radiation is the first correction in a non-relativistic velocity expansion and dipole emission dominates. 
For example, if one wishes to compute the Gaunt factor for electron-electron bremsstrahlung one uses Eq.~\eqref{eq:bornnonid} for $d\sigma/dx|_{\substack{\text{\tiny Born}\\ \text{\tiny non-id}}}$ with $Z_{1,2}=1$, $m_{1,2}=m_e$ and $\mu = m_e/2$ together with the tabulations of $\gnonid$ and $\gint$ for repulsive interactions.
If one wishes to compute the quadrupole cross section for electron-ion bremsstrahlung, one simply uses Eq.~\eqref{eq:bornnonid} for $d\sigma/dx|_{\substack{\text{\tiny Born}\\ \text{\tiny non-id}}}$ (with the appropriate mass and charge assignments) in the defining equation~\eqref{eq:GauntQuadrupole} together with the tabulation of $\gnonid$ for attractive interactions.

The Elwert factor $e^{-2\pi \nu_i} S_f/S_i$ in Eq.~\eqref{eq:GauntQuadrupole} approaches unity for $x\ll1$ and enhances (suppresses) the cross section with respect to the Born limit at the kinematic endpoint for attractive (repulsive) interactions.
Therefore, the exact cross section in Eq.~\eqref{eq:GauntQuadrupole} is always suppressed with respect to the Born limit for repulsive interactions. 

In the classical limit, the \lss{exchange} term that is present for repulsive interactions tends to zero regardless of the symmetry properties of the colliding particles,
\begin{align}
 \gint \to 0 \quad \text{for}\quad  |\nu_i|\to \infty .
\end{align}
\lss{In the same limit $\gnonid$ tends to zero for $|x\nu_i| \to 0$ and diverges for $|x\nu_i| \to \infty$. Of course, the  cross section itself remains finite~\citep{Pradler:2020znn},
\begin{align}
   x \frac{d\sigma_{Q}}{dx} &=
   \frac{8\alpha^3 Z_1^2 Z_2^2 \mu^2}{15}
		\left(\frac{Z_1}{m_1^2} + \frac{Z_2}{m_2^2}\right)^2 \nonumber \\ & \!\!\!\!\times
   \begin{cases}\left[
	10 + 6 \ln\left(\frac{4  e^{-\gamma}}{x \, |\nu_i|}\right) \right]
	& \text{for}\,\, |\nu_i|\gg 1 \land   |x \nu_i| \ll 1, \\
	\frac{60 \pi^{3/2}}{3^{7/6} \Gamma(1/6)} \left(\frac{x \, |\nu_i|}{2}\right)^{2/3}
	& \text{for}\,\, |\nu_i|\gg 1 \land |x \nu_i| \gg 1,
   \end{cases}
\end{align}
where the bottom expression in the second line is valid for attractive interactions ($Z_1 Z_2 < 0$) and is to be multiplied by $\exp(-\pi x |\nu_i|)$ for repulsive interactions ($Z_1 Z_2 > 0$). 
% with
% \begin{align}
% x \frac{d\sigma_{Q}}{dx} &=
%     \frac{8\alpha^3 Z_1^2 Z_2^2 \mu^2}{15}
% 		\left(\frac{Z_1}{m_1^2} + \frac{Z_2}{m_2^2}\right)^2 \times G .
% \end{align}
Finally,} in the Born limit, the Gaunt factor for non-identical particle scattering approaches unity, $\gnonid \to 1$, and
\begin{align}
 \gint  \to 0 (1) \quad \text{for}\quad \ |\nu_i|\to 0,\ x\to 0(1) .
\end{align}
Generally, $ \gint < 1$ away from the kinematic endpoint; the limit for $x\to 0$ is only approached logarithmically as $\gint \sim 3/[10+6\ln(4/x)]$. See App.~\ref{app:born_cs} for the expression of $\gint $ in the Born limit.

\subsection{Numerical evaluation}
\label{sec:num_gaunt}

We now describe the numerical evaluation of the Gaunt factor across a wide range of the parameters $\nu_i$ and $x$.
The integrand of the exact double differential cross sections in $x$ and scattering angle $\cos\theta$ is obtained by using a dedicated C library for arbitrary-precision ball arithmetic~\texttt{Arb}~\citep{Johansson2017arb}. A particular strength of it is the rigorous computation of hypergeometric functions~\citep{2016arXiv160606977J};
common software frameworks such as \texttt{Mathematica} appear not able to yield results for $|\nu_i|\gtrsim 100$ for all required function values in a reasonable amount of time, let alone to integrate them to a single differential cross section $d\sigma/dx$ at high precision. Whenever the evaluation is possible with both software products, we have verified that their results agree. 

Concretely, $d\sigma/(dx\, d\cos\theta)$ is given  by~Eqs.~(34) and (39) in \cite{Pradler:2020znn}. We use \texttt{Arb} to calculate the hypergeometric functions, Sommerfeld factors and all coefficients; in short, the entire expression. This is required, because, first, products such as $S_i S_f \times |{}_2F_1(i\nu_f,i\nu_i;1;z) |^2$ can be of the form ``huge'' $\times$ ``tiny'' and, second,  because of the occurrence of cancellations in the final linear combination of these terms. Because of factors $\exp(\pm \nu_{i,f})$ contained in $S_{i,f}$, it turns out that the required precision is approximately $|\nu_{i,f}|$. In our tabulation we therefore evaluate the ingredients with up to $10^4$ digits of precision.

To obtain the Gaunt factors, we convert the final result for the integrand to double precision and then perform the angular integration using an adaptive integration routine (61-point Gauss-Kronrod rule) of the GNU Scientific Library \citep{galassi2018scientific}. The results for $\gnonid$ and $\gint$ are shown in the left and right panel of Fig.~\ref{fig:g_ff} respectively.
In addition, we present the values of the Gaunt factors with 6 significant digits in four supplementary electronic tables.  We provide $\gnonid$ and $\gint$ on a logarithmic grid in $x$ for $-12-\epsilon \leq \log_{10}x \leq-\epsilon$ with $\epsilon=10^{-4}$ and a spacing of $0.025 \, \text{dex}$ in Tab.~\ref{tab:1} and, in addition, on a linear grid for $0.0025 \leq x \leq 0.9975$ with a spacing of $\Delta x=0.0025$ in Tab.~\ref{tab:6} in App.~\ref{app:addtables}. The linear grid better resolves the oscillations of $\gint$ that can be seen in the bottom panel of Fig.~\ref{fig:g_ff}.
In both cases, we have used a logarithmic grid in $\nu_i$ for $-3 \leq \log_{10}|\nu_i| \leq 4$  and a spacing of $0.1$~dex. For $|\nu_i| \geq 10$, we set $\gint=0$ since its relative importance in Eq.~\eqref{eq:gaunt_id} at that point has dropped several orders of magnitude below the precision presented here. 
% \lss{The grid is fine enough such that a bi-linear interpolation of the electronic tables should suffice for all practical purposes.}
As a service, we also provide the differential electron-electron bremsstrahlung cross section $\omega d\sigma_{Q}/d\omega$ in units of barn on the double logarithmic grid in initial kinetic energy $m_e v_i^2/4$ and emitted photon energy $\omega$ in Tab.~\ref{tab:gee}.
\lss{When interpolating the tables provided in this section to calculate values of $\gid$ and $\gnonid$ away from the grid points, we have checked that a bi-linear interpolation in $\log x$ and $\log \nu_i$ yields a maximum error of roughly $10^{-3}$ at the kinematic endpoint $x\sim 1$. 
Using a bi-cubic interpolation, the relative error stays below $10^{-6}$ across the entire kinematic regime.
}

\begin{table*}[tp]
\centering
\setlength{\tabcolsep}{6.5pt}
\renewcommand{\arraystretch}{\tablesepvert}
\begin{tabular}{r | rrrrrrrrrrr } 
\multicolumn{12}{c}{Gaunt factors $\gnonid$ (top) and $\gint$ (bottom)} \\[3pt]
 \toprule
  \multicolumn{1}{c|}{}&\multicolumn{11}{c}{$\log_{10}|\nu_i|$} \\
  \multicolumn{1}{c|}{$\log_{10}x$} &
  \multicolumn{1}{c}{\Minus1.0} & \multicolumn{1}{c}{\Minus0.5} & \multicolumn{1}{c}{0.0} & \multicolumn{1}{c}{0.5} & \multicolumn{1}{c}{1.0} & \multicolumn{1}{c}{1.5} & \multicolumn{1}{c}{2.0} & \multicolumn{1}{c}{2.5} & \multicolumn{1}{c}{3.0} & \multicolumn{1}{c}{3.5} & \multicolumn{1}{c}{4.0}    \\
 \midrule
\Minus12.0001 & 1.00\Plus0 & 9.96\Minus1 & 9.78\Minus1 & 9.43\Minus1 & 9.06\Minus1 & 8.69\Minus1 & 8.31\Minus1 & 7.94\Minus1 & 7.56\Minus1 & 7.19\Minus1 & 6.81\Minus1 \\
\Minus11.0001 & 1.00\Plus0 & 9.96\Minus1 & 9.76\Minus1 & 9.39\Minus1 & 8.99\Minus1 & 8.58\Minus1 & 8.17\Minus1 & 7.77\Minus1 & 7.36\Minus1 & 6.96\Minus1 & 6.55\Minus1 \\
\Minus10.0001 & 1.00\Plus0 & 9.96\Minus1 & 9.74\Minus1 & 9.33\Minus1 & 8.90\Minus1 & 8.45\Minus1 & 8.01\Minus1 & 7.57\Minus1 & 7.13\Minus1 & 6.69\Minus1 & 6.25\Minus1 \\
\Minus9.0001 & 9.99\Minus1 & 9.95\Minus1 & 9.72\Minus1 & 9.27\Minus1 & 8.79\Minus1 & 8.30\Minus1 & 7.82\Minus1 & 7.34\Minus1 & 6.85\Minus1 & 6.37\Minus1 & 5.88\Minus1 \\
\Minus8.0001 & 9.99\Minus1 & 9.95\Minus1 & 9.69\Minus1 & 9.19\Minus1 & 8.66\Minus1 & 8.12\Minus1 & 7.59\Minus1 & 7.05\Minus1 & 6.51\Minus1 & 5.98\Minus1 & 5.44\Minus1 \\
\Minus7.0001 & 9.99\Minus1 & 9.94\Minus1 & 9.65\Minus1 & 9.09\Minus1 & 8.50\Minus1 & 7.90\Minus1 & 7.30\Minus1 & 6.70\Minus1 & 6.10\Minus1 & 5.50\Minus1 & 4.90\Minus1 \\
\Minus6.0001 & 9.99\Minus1 & 9.93\Minus1 & 9.60\Minus1 & 8.97\Minus1 & 8.29\Minus1 & 7.61\Minus1 & 6.93\Minus1 & 6.25\Minus1 & 5.57\Minus1 & 4.90\Minus1 & 4.26\Minus1 \\
\Minus5.0001 & 9.99\Minus1 & 9.92\Minus1 & 9.54\Minus1 & 8.81\Minus1 & 8.02\Minus1 & 7.24\Minus1 & 6.45\Minus1 & 5.68\Minus1 & 4.93\Minus1 & 4.26\Minus1 & 3.73\Minus1 \\
\Minus4.0001 & 9.99\Minus1 & 9.91\Minus1 & 9.45\Minus1 & 8.59\Minus1 & 7.66\Minus1 & 6.75\Minus1 & 5.86\Minus1 & 5.06\Minus1 & 4.43\Minus1 & 4.14\Minus1 & 4.45\Minus1 \\
\Minus3.0001 & 9.99\Minus1 & 9.89\Minus1 & 9.34\Minus1 & 8.29\Minus1 & 7.21\Minus1 & 6.22\Minus1 & 5.45\Minus1 & 5.10\Minus1 & 5.48\Minus1 & 7.17\Minus1 & 1.14\Plus0 \\
\Minus2.0001 & 9.99\Minus1 & 9.87\Minus1 & 9.21\Minus1 & 8.05\Minus1 & 7.06\Minus1 & 6.60\Minus1 & 7.11\Minus1 & 9.31\Minus1 & 1.48\Plus0 & 2.69\Plus0 & 5.32\Plus0 \\
\Minus1.0001 & 9.99\Minus1 & 9.91\Minus1 & 9.48\Minus1 & 9.07\Minus1 & 9.89\Minus1 & 1.31\Plus0 & 2.10\Plus0 & 3.85\Plus0 & 7.64\Plus0 & 1.58\Plus1 & 3.35\Plus1 \\
\Minus0.0001 & 1.00\Plus0 & 1.04\Plus0 & 1.29\Plus0 & 2.01\Plus0 & 3.63\Plus0 & 7.13\Plus0 & 1.47\Plus1 & 3.10\Plus1 & 6.61\Plus1 & 1.42\Plus2 & 3.05\Plus2 \\
 \midrule
\Minus12.0001 & 1.60\Minus2 & 1.39\Minus2 & 4.43\Minus3 & 1.57\Minus5 & \multicolumn{1}{c}{\dots} & \multicolumn{1}{c}{\dots} & \multicolumn{1}{c}{\dots} & \multicolumn{1}{c}{\dots} & \multicolumn{1}{c}{\dots} & \multicolumn{1}{c}{\dots} & \multicolumn{1}{c}{\dots} \\
\Minus11.0001 & 1.73\Minus2 & 1.50\Minus2 & 4.79\Minus3 & 1.70\Minus5 & \multicolumn{1}{c}{\dots} & \multicolumn{1}{c}{\dots} & \multicolumn{1}{c}{\dots} & \multicolumn{1}{c}{\dots} & \multicolumn{1}{c}{\dots} & \multicolumn{1}{c}{\dots} & \multicolumn{1}{c}{\dots} \\
\Minus10.0001 & 1.89\Minus2 & 1.63\Minus2 & 5.22\Minus3 & 1.85\Minus5 & \multicolumn{1}{c}{\dots} & \multicolumn{1}{c}{\dots} & \multicolumn{1}{c}{\dots} & \multicolumn{1}{c}{\dots} & \multicolumn{1}{c}{\dots} & \multicolumn{1}{c}{\dots} & \multicolumn{1}{c}{\dots} \\
\Minus9.0001 & 2.07\Minus2 & 1.79\Minus2 & 5.72\Minus3 & 2.03\Minus5 & \multicolumn{1}{c}{\dots} & \multicolumn{1}{c}{\dots} & \multicolumn{1}{c}{\dots} & \multicolumn{1}{c}{\dots} & \multicolumn{1}{c}{\dots} & \multicolumn{1}{c}{\dots} & \multicolumn{1}{c}{\dots} \\
\Minus8.0001 & 2.29\Minus2 & 1.99\Minus2 & 6.33\Minus3 & 2.24\Minus5 & \multicolumn{1}{c}{\dots} & \multicolumn{1}{c}{\dots} & \multicolumn{1}{c}{\dots} & \multicolumn{1}{c}{\dots} & \multicolumn{1}{c}{\dots} & \multicolumn{1}{c}{\dots} & \multicolumn{1}{c}{\dots} \\
\Minus7.0001 & 2.57\Minus2 & 2.22\Minus2 & 7.09\Minus3 & 2.51\Minus5 & \multicolumn{1}{c}{\dots} & \multicolumn{1}{c}{\dots} & \multicolumn{1}{c}{\dots} & \multicolumn{1}{c}{\dots} & \multicolumn{1}{c}{\dots} & \multicolumn{1}{c}{\dots} & \multicolumn{1}{c}{\dots} \\
\Minus6.0001 & 2.92\Minus2 & 2.53\Minus2 & 8.06\Minus3 & 2.85\Minus5 & \multicolumn{1}{c}{\dots} & \multicolumn{1}{c}{\dots} & \multicolumn{1}{c}{\dots} & \multicolumn{1}{c}{\dots} & \multicolumn{1}{c}{\dots} & \multicolumn{1}{c}{\dots} & \multicolumn{1}{c}{\dots} \\
\Minus5.0001 & 3.38\Minus2 & 2.93\Minus2 & 9.34\Minus3 & 3.31\Minus5 & \multicolumn{1}{c}{\dots} & \multicolumn{1}{c}{\dots} & \multicolumn{1}{c}{\dots} & \multicolumn{1}{c}{\dots} & \multicolumn{1}{c}{\dots} & \multicolumn{1}{c}{\dots} & \multicolumn{1}{c}{\dots} \\
\Minus4.0001 & 4.01\Minus2 & 3.48\Minus2 & 1.11\Minus2 & 3.93\Minus5 & \multicolumn{1}{c}{\dots} & \multicolumn{1}{c}{\dots} & \multicolumn{1}{c}{\dots} & \multicolumn{1}{c}{\dots} & \multicolumn{1}{c}{\dots} & \multicolumn{1}{c}{\dots} & \multicolumn{1}{c}{\dots} \\
\Minus3.0001 & 4.94\Minus2 & 4.28\Minus2 & 1.37\Minus2 & 4.84\Minus5 & \multicolumn{1}{c}{\dots} & \multicolumn{1}{c}{\dots} & \multicolumn{1}{c}{\dots} & \multicolumn{1}{c}{\dots} & \multicolumn{1}{c}{\dots} & \multicolumn{1}{c}{\dots} & \multicolumn{1}{c}{\dots} \\
\Minus2.0001 & 6.43\Minus2 & 5.57\Minus2 & 1.78\Minus2 & 6.38\Minus5 & \multicolumn{1}{c}{\dots} & \multicolumn{1}{c}{\dots} & \multicolumn{1}{c}{\dots} & \multicolumn{1}{c}{\dots} & \multicolumn{1}{c}{\dots} & \multicolumn{1}{c}{\dots} & \multicolumn{1}{c}{\dots} \\
\Minus1.0001 & 9.51\Minus2 & 8.16\Minus2 & 2.41\Minus2 & 4.43\Minus5 & \multicolumn{1}{c}{\dots} & \multicolumn{1}{c}{\dots} & \multicolumn{1}{c}{\dots} & \multicolumn{1}{c}{\dots} & \multicolumn{1}{c}{\dots} & \multicolumn{1}{c}{\dots} & \multicolumn{1}{c}{\dots} \\
\Minus0.0001 & 9.94\Minus1 & 9.47\Minus1 & 5.63\Minus1 & 6.46\Minus3 & \multicolumn{1}{c}{\dots} & \multicolumn{1}{c}{\dots} & \multicolumn{1}{c}{\dots} & \multicolumn{1}{c}{\dots} & \multicolumn{1}{c}{\dots} & \multicolumn{1}{c}{\dots} & \multicolumn{1}{c}{\dots} \\

 \bottomrule
\end{tabular}
\tablecomments{\TableComment}
\caption{Free-free Gaunt factor for non-identical particles $\gnonid$ (top) and the \lss{exchange} term for identical particles $\gint$ (bottom) as defined through Eqs.~\eqref{eq:GauntQuadrupole} and~\eqref{eq:gaunt_id} in the range $10^{-3}\le|\nu_i|\le 10^4$ in increments of 0.1~dex and $10^{-12-\epsilon}<x<10^{-\epsilon}$ with $\epsilon=10^{-4}$ on a logarithmic grid in $x$ in increments of 0.025~dex. The values are computed using Eq.~(34) in \cite{Pradler:2020znn}; the notation $1.60\!-\!2$ means a value of $1.60\times 10^{-2}$ and so forth. 
For $|\nu_i|\geq 10$
 we set $\gint=0$, indicated by the dots, as their magnitude is smaller than the quoted precision.
\label{tab:1}}
\end{table*}

\begin{table*}[tp]
\centering
\setlength{\tabcolsep}{5.5pt}
\renewcommand{\arraystretch}{\tablesepvert}
\begin{tabular}{r | lllllllllll } 
\multicolumn{12}{c}{Electron-electron cross section $\omega d\sigma_{Q}/d\omega$ (barn)} \\[3pt]
 \toprule
  \multicolumn{1}{c|}{}&\multicolumn{11}{c}{$\log_{10}\left[(m_e v_i^2/4)/\mathrm{eV}\right]$} \\
  \multicolumn{1}{c|}{$\log_{10}(\omega/\mathrm{eV})$} &
  \multicolumn{1}{c}{\Minus6.0} & \multicolumn{1}{c}{\Minus5.0} & \multicolumn{1}{c}{\Minus4.0} &\multicolumn{1}{c}{\Minus3.0} &\multicolumn{1}{c}{\Minus2.0} &\multicolumn{1}{c}{\Minus1.0} &  \multicolumn{1}{c}{0.0} & \multicolumn{1}{c}{1.0} & \multicolumn{1}{c}{2.0} & \multicolumn{1}{c}{3.0} & \multicolumn{1}{c}{4.0} \\
 \midrule
\Minus8.0 & 8.39\Minus37 & 7.49\Minus4 & 1.09\Minus2 & 1.77\Minus2 & 2.42\Minus2 & 3.06\Minus2 & 3.69\Minus2 & 4.29\Minus2 & 4.78\Minus2 & 5.22\Minus2 & 5.64\Minus2 \\
\Minus7.0 & \multicolumn{1}{c}{---} & 1.90\Minus13 & 4.16\Minus3 & 1.33\Minus2 & 1.99\Minus2 & 2.63\Minus2 & 3.27\Minus2 & 3.87\Minus2 & 4.35\Minus2 & 4.79\Minus2 & 5.22\Minus2 \\
\Minus6.0 & \multicolumn{1}{c}{---} & 3.30\Minus119 & 3.73\Minus6 & 7.92\Minus3 & 1.56\Minus2 & 2.20\Minus2 & 2.84\Minus2 & 3.44\Minus2 & 3.93\Minus2 & 4.36\Minus2 & 4.79\Minus2 \\
\Minus5.0 & \multicolumn{1}{c}{---} & \multicolumn{1}{c}{---} & 2.38\Minus39 & 7.33\Minus4 & 1.09\Minus2 & 1.77\Minus2 & 2.41\Minus2 & 3.01\Minus2 & 3.50\Minus2 & 3.93\Minus2 & 4.36\Minus2 \\
\Minus4.0 & \multicolumn{1}{c}{---} & \multicolumn{1}{c}{---} & \multicolumn{1}{c}{---} & 2.94\Minus14 & 4.12\Minus3 & 1.33\Minus2 & 1.99\Minus2 & 2.58\Minus2 & 3.07\Minus2 & 3.51\Minus2 & 3.94\Minus2 \\
\Minus3.0 & \multicolumn{1}{c}{---} & \multicolumn{1}{c}{---} & \multicolumn{1}{c}{---} & \multicolumn{1}{c}{---} & 2.01\Minus6 & 7.86\Minus3 & 1.55\Minus2 & 2.16\Minus2 & 2.64\Minus2 & 3.08\Minus2 & 3.51\Minus2 \\
\Minus2.0 & \multicolumn{1}{c}{---} & \multicolumn{1}{c}{---} & \multicolumn{1}{c}{---} & \multicolumn{1}{c}{---} & \multicolumn{1}{c}{---} & 5.83\Minus4 & 1.08\Minus2 & 1.73\Minus2 & 2.22\Minus2 & 2.65\Minus2 & 3.08\Minus2 \\
\Minus1.0 & \multicolumn{1}{c}{---} & \multicolumn{1}{c}{---} & \multicolumn{1}{c}{---} & \multicolumn{1}{c}{---} & \multicolumn{1}{c}{---} & \multicolumn{1}{c}{---} & 3.68\Minus3 & 1.28\Minus2 & 1.79\Minus2 & 2.23\Minus2 & 2.65\Minus2 \\
0.0 & \multicolumn{1}{c}{---} & \multicolumn{1}{c}{---} & \multicolumn{1}{c}{---} & \multicolumn{1}{c}{---} & \multicolumn{1}{c}{---} & \multicolumn{1}{c}{---} & \multicolumn{1}{c}{---} & 6.99\Minus3 & 1.35\Minus2 & 1.80\Minus2 & 2.23\Minus2 \\
1.0 & \multicolumn{1}{c}{---} & \multicolumn{1}{c}{---} & \multicolumn{1}{c}{---} & \multicolumn{1}{c}{---} & \multicolumn{1}{c}{---} & \multicolumn{1}{c}{---} & \multicolumn{1}{c}{---} & \multicolumn{1}{c}{---} & 8.39\Minus3 & 1.36\Minus2 & 1.80\Minus2 \\
2.0 & \multicolumn{1}{c}{---} & \multicolumn{1}{c}{---} & \multicolumn{1}{c}{---} & \multicolumn{1}{c}{---} & \multicolumn{1}{c}{---} & \multicolumn{1}{c}{---} & \multicolumn{1}{c}{---} & \multicolumn{1}{c}{---} & \multicolumn{1}{c}{---} & 8.75\Minus3 & 1.37\Minus2 \\
3.0 & \multicolumn{1}{c}{---} & \multicolumn{1}{c}{---} & \multicolumn{1}{c}{---} & \multicolumn{1}{c}{---} & \multicolumn{1}{c}{---} & \multicolumn{1}{c}{---} & \multicolumn{1}{c}{---} & \multicolumn{1}{c}{---} & \multicolumn{1}{c}{---} & \multicolumn{1}{c}{---} & 8.85\Minus3 \\
4.0 & \multicolumn{1}{c}{---} & \multicolumn{1}{c}{---} & \multicolumn{1}{c}{---} & \multicolumn{1}{c}{---} & \multicolumn{1}{c}{---} & \multicolumn{1}{c}{---} & \multicolumn{1}{c}{---} & \multicolumn{1}{c}{---} & \multicolumn{1}{c}{---} & \multicolumn{1}{c}{---} & \multicolumn{1}{c}{---} \\

 \bottomrule
\end{tabular}
\tablecomments{\TableComment}
\caption{Free-free differential cross section $\omega d\sigma_{Q}/d\omega$ for electron-electron bremsstrahlung in the range $0.1\,\mu\text{eV}\le m_e v_i^2/4 \le 10 \, \text{keV}$ in increments of 0.1~dex and $10\,\text{neV}<\omega<10 \, \text{keV}$ in increments of 0.05~dex. The values are computed using Eq.~(39) in \cite{Pradler:2020znn}. Dashes indicate  values that are outside the kinematically allowed range or are too small to be represented double precision numbers and are set to zero in the electronic version of the table; the notation follows Tab.~\ref{tab:1}.
\label{tab:gee}}
\end{table*}

\subsection{Approximate bremsstrahlung formula}
In the soft-photon limit $x\ll 1$, \cite{Weinberg:2019mai} recently showed how an accurate dipole radiation formula can be obtained that is valid for arbitrary $|\nu_i|$. Building on those insights, in~\cite{Pradler:2020znn} an analogous formula was established for soft quadrupole radiation and which reproduced in the appendix in~\eqref{eqn:quad_emission_weinberg}. It likewise has the correct asymptotic forms in the Born and the classical limits.
For repulsive interactions, it is actually possible to go beyond the soft-photon limit and obtain an approximate formula across the entire range of parameters, and which is numerically much easier to evaluate than the exact expression~\citep{Pradler:2020znn},
\begin{widetext}
\begin{align} \label{eqn:quad_approximate}
   \left. x \frac{d \sigma_{Q}}{dx}\right|_{\rm approx} &\simeq
    \frac{4\alpha^{3}Z_1^2 Z_2^2  \mu^2 }{15}
   \left(\frac{Z_1}{m_1^2} +\frac{Z_2}{m_2^2} \right)^2
    \frac{\sqrt{1-x}}{e^{\pi x |\nu_i|}}
    \Bigg\{
    \frac{40 \pi^{3/2} \left[\frac{x|\nu_i|}{2}\right]^{\frac{2}{3}}}{3^{1/6} \Gamma(1/6)} 
    + \frac{\pi^2 \nu_i^2}{\sinh^2{\pi \nu_i}}\left[6\frac{2+\tilde\delta_{12}}{\zeta^2+1}-3(2+\tilde\delta_{12})+ 12\ln \zeta
    \right] 
    \nonumber \\
    &
    +26 
    -6\frac{2+\tilde\delta_{12}}{\zeta^2+1}
    +12\ln \left[\frac{1}{\zeta}\frac{1+\sqrt{1-x}}{1-\sqrt{1-x}}\right] 
    \Bigg\} 
    ,
\end{align}
\end{widetext}
where $\zeta=|\nu_i| e^{\gamma+1/2}$, $\gamma = 0.5772\dots$ is the Euler-Mascheroni constant, and $\tilde\delta_{12}=0$ for non-identical particles and $\tilde\delta_{12}=1(-2)$ for identical  spin-1/2 (spin-0) particles. 
The validity of~\eqref{eqn:quad_approximate} as function of $x$ and $\nu_i$ in its approximation to the exact result has been studied in~\cite{Pradler:2020znn}.
In the following sections that deal with thermal averaged versions of the Gaunt factor, we shall also obtain approximate versions through the use of Eq.~\eqref{eqn:quad_approximate} and compare the results.

\section{Thermally Averaged Gaunt Factor}
\label{sec:therm_gaunt}

\begin{figure*}[tb]
		\includegraphics[width=\textwidth]{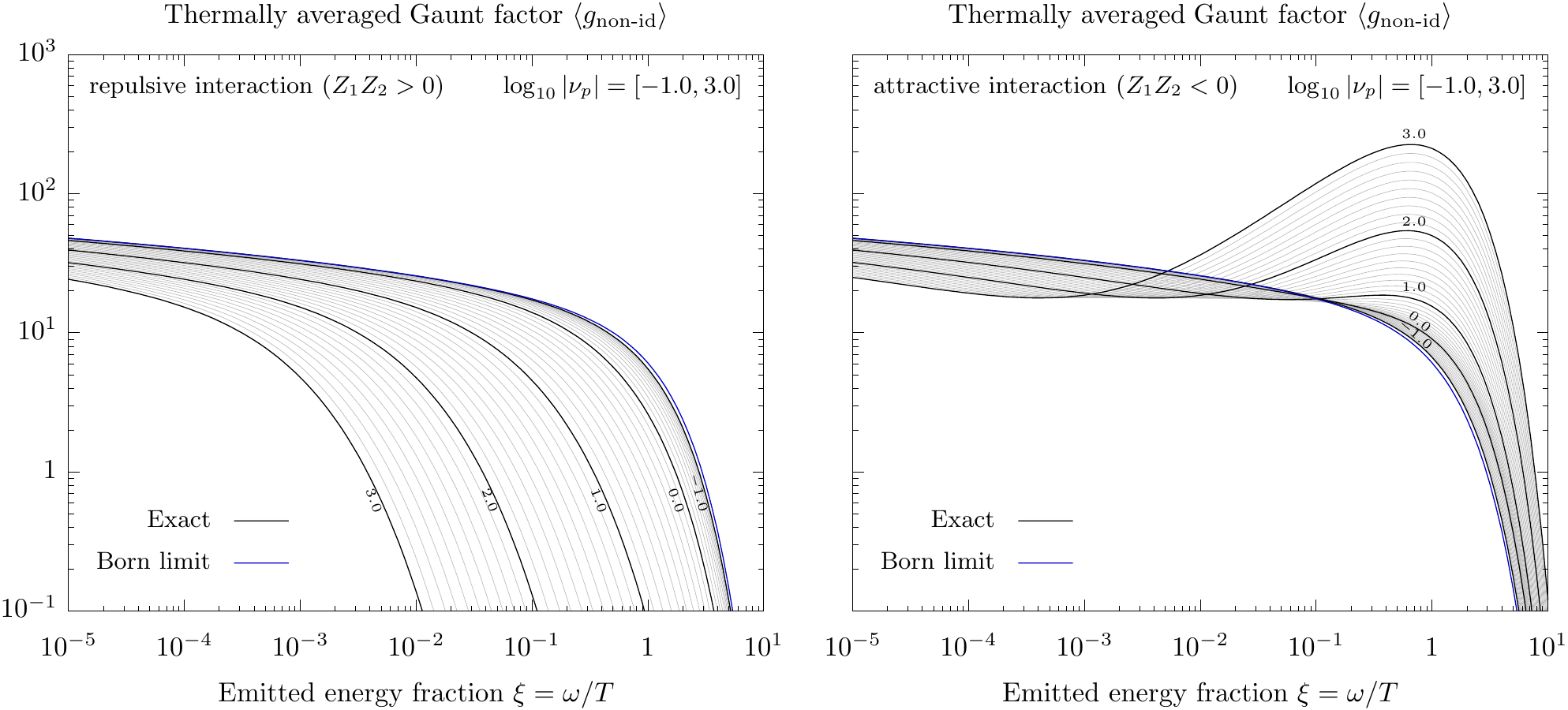}
		\caption{Thermally averaged free-free Gaunt factor for non-identical particles for repulsive interactions (left panel), and for the attractive interactions (right panel) over a wide range of parameters $\nu_p$ and $\xi$. The lines  are shown in increments of $\log_{10}(|\nu_p|)=0.1$ and thicker lines are for $\nu_p$ values as labeled. Born limits are shown in blue.  \label{fig:gth_ff}}
\end{figure*}

\begin{figure}[tb]
		\includegraphics[width=\columnwidth]{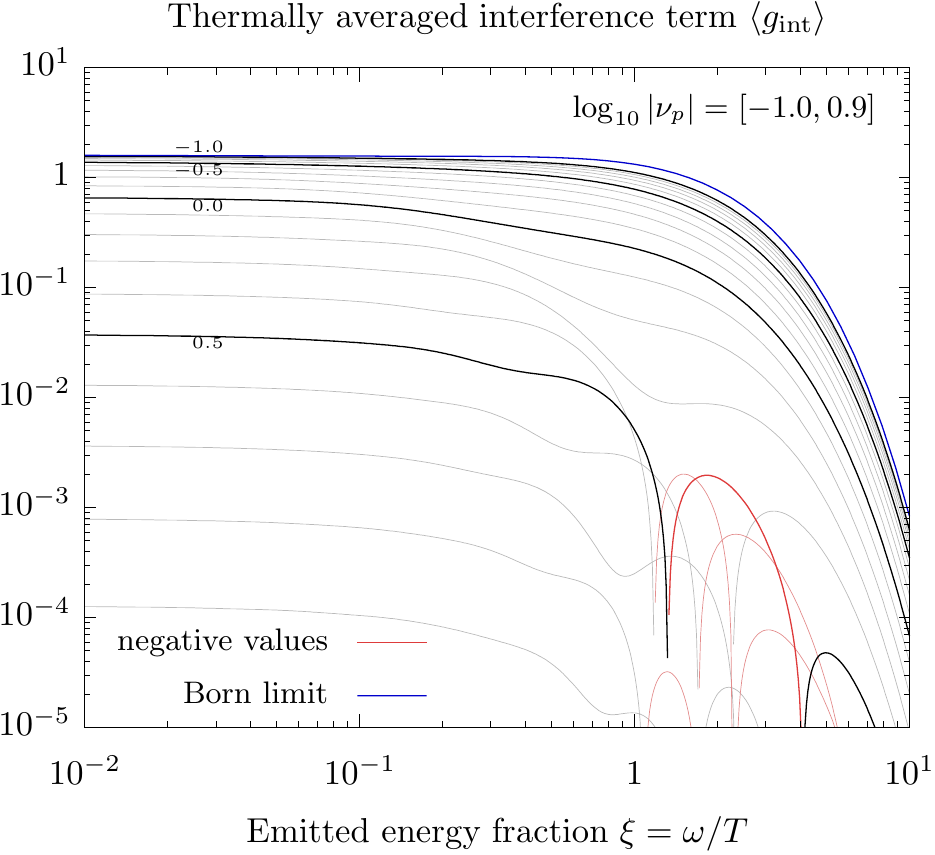}
		\caption{Thermally averaged free-free \lss{exchange}  term over a wide range of parameters $\nu_p$ and $\xi$. The lines  are shown in increments of $\log_{10}(|\nu_p|)=0.1$ and thicker lines are for $\nu_p$ values as labeled. Negative values are shown by the red line-segments; Born limits are shown in blue.  \label{fig:gth_int}}
\end{figure}

The dynamics of charged particles and photons in a plasma is rather complicated.
However, for many astrophysical applications one can neglect in-medium effects such as degeneracy, screening or a finite photon mass, and treat the plasma as a non-degenerate one.
Concretely, when $\omega \gg v_i k_D$ and $\omega \gg \omega_p$ where $k_D$ and $\omega_p$ are the Debye scale and plasma frequency, respectively, medium effects can be neglected in the bremsstrahlung emission. In the post-recombination dilute interstellar medium this is both satisfied for $\omega > 10^{-10}\,{\rm eV}$ or, equivalently, for frequencies in excess of $24$~kHz. We may hence take a simple thermal average of the Gaunt factor  while retaining broad applicability; for a detailed discussion on medium effects in this context, see Sec.~9 in \cite{Pradler:2020znn}.

\subsection{Free-Free emission}
\label{sec:def_gaunt_th}

Following tradition~\citep{1961ApJS....6..167K}, we introduce the dimensionless energy $\xi$ and  the most probable value of $|\nu_i|$, $|\nu_p|$, as well as the dimensionless incoming CM energy $u$ in units of common plasma temperature~$T$, 
\begin{align}
    \xi\equiv \frac{\omega}{T},
    \quad \nu_p \equiv - Z_1 Z_2 \alpha \sqrt{\frac{\mu}{2T}}, \quad u \equiv \frac{\mu v_i^2}{2T},
\end{align}
respectively.
The thermally averaged emission cross section in a non-relativistic Maxwellian plasma then reads,
{
\medmuskip=1mu
\thinmuskip=1mu
\thickmuskip=1mu
\nulldelimiterspace=1pt
\scriptspace=1pt
\begin{equation}
 \label{eq:ThermSigma}
    d\left\langle \sigma_Q v \right\rangle (\xi, \nu_p)  = \sqrt{\frac{8}{\pi}\frac{T}{\mu}} \int_\xi^\infty du \, u\,
   e^{-u}\;  d\sigma_Q\left(x = \frac{\xi}{u}  , \nu_i = \frac{\nu_p}{\sqrt{u}}\right),
\end{equation}
}%
giving rise to the definition of thermally averaged free-free Gaunt factor as%
\begin{align} \label{eq:def_therm_avg_gaunt}
    \xi \frac{ d\left\langle \sigma_Q v \right\rangle }{d\xi} (\xi, \nu_p) =
     \sqrt{\frac{8}{\pi}\frac{T}{\mu}}
    \frac{\alpha^3 Z_1^2 Z_2^2 A_Q^2}{\mu^2} 
     \langle  \gff  \rangle (\xi, \nu_p) .
\end{align}
with $\langle\gid\rangle = \langle\gnonid\rangle + (-1)^{2s}\langle\gint\rangle/(2s+1) $ and where $A_Q = \mu^2 (Z_1/m_1^2 + Z_2/m_2^2)$; $A_Q=1/2$ for electron-electron and $A_Q\simeq 1$ for electron-ion scattering.
Temperatures of primary astrophysical interest range from $T\sim 1 \,\text{eV}$ found \eg~in HII regions to $T\sim 10 \,\text{keV}$ in a hot intracluster gas. Values of $10^{-2}<|\nu_p|<10^3$ cover these temperatures for the mutual scattering of electrons, for electron-ion scattering, as well as the mutual scattering of ionized hydrogen, or fully ionized helium. Note that the symmetry property of the Gaunt factor
under the sign change of $\nu_i$ is not preserved in the thermal average, $\langle\gff\rangle(\xi, \nu_p)\neq\langle\gff\rangle(\xi, -\nu_p)$, since the definition of $\langle\gff\rangle$ in Eq.~\eqref{eq:def_therm_avg_gaunt} includes the Elwert factor which is greater (less) than unity for positive (negative) $\nu_p$ in the thermal average.

The averaged Gaunt factor can be used to calculate the production spectrum of photons, \ie, the production rate of photons per volume and photon energy, due to bremsstrahlung in a non-relativistic plasma of particles with number densities $n_1$ and $n_2$ (see App.~\ref{app:thermal} for details), 
\begin{align} \label{eq:spectrum}
  \frac{d\Gamma^{(Q)}_{\rm brem}}{dV d\omega}  
  &=
  \sqrt{\frac{8}{\pi} \frac{T}{\mu^5}}
  \frac{n_{1} n_{2}}{1+\delta_{12}}
  \frac{\alpha^3 Z_1^2 Z_2^2 A_Q^2}{\omega}  \langle  \gff  \rangle , 
\end{align}
with $\delta_{12}= 1(0)$ for (non-)identical particles.
 The energy spectrum of emitted photons is trivially obtained from Eq.~\eqref{eq:spectrum} by multiplying with $\omega$.
Finally, we recall that the emissivity $j_\nu(T)$ is the emitted energy at temperature $T$ and at a specific frequency~$\nu = 2\pi \omega$ per time, per volume, per solid angle, and per frequency interval. Specialized to electrons, $n_{1,2}=n_e$, and using  ordinary units to make better contact with the astrophysics literature, it is given by
\begin{subequations}
\begin{align}
    j^{(Q)}_\nu(T) &= \frac{1}{4\pi} h\nu \frac{n_e^2}{2} \frac{d\langle \sigma_Q v \rangle}{d\nu} \\ 
    % & \simeq 1.34\times 10^{-45}\,n_e^2 \, T_8^{1/2} \langle \gff \rangle\: \frac{\rm erg\, cm^3}{\rm  s\, Hz\, sr}
    & \simeq 1.34\times 10^{-45}\, T_8^{1/2} \lss{\left(\frac{n_e}{\rm cm^3}\right)^2}   \langle \gff \rangle\: \lss{\frac{\rm erg}{\rm cm^3\, s\, Hz\, sr}}
\end{align}
\end{subequations}
In the second line $T_8 = T/10^8\,\rm K$ and  the Gaunt factor is to be evaluated at $\xi=3.02\times 10^{-9} \nu_{\rm GHz}T_8^{-1} $ and $ |\nu_p|= 0.028 T_8^{-1/2}$ where $\nu_{\rm GHz} = \nu/10^{9}\,\rm Hz$.

\subsection{Free-Free absorption}
\label{sec:abs}

The relation between bremsstrahlung emission and photon absorption (restricted to quadrupole transitions) in a medium of joint temperature $T$ can be established from the Boltzmann-type equation,
\begin{align}
    \frac{dn_\gamma}{d\omega dt} = \frac{d\Gamma^{(Q)}_\text{brem}}{d\omega dV} \frac{1}{1-e^{-\omega/T}} - \frac{dn_\gamma}{d\omega} \Gamma^{(Q)}_{\rm abs},
\end{align}
where $n_\gamma$ is the number density of photons in the interval $(\omega, \omega+d\omega)$ and $\Gamma^{(Q)}_{\text{abs}}$ is the total quadrupole photon absorption rate for a photon of energy $\omega$; the factor multiplying ${d\Gamma^{(Q)}_\text{brems}}$ corrects for stimulated emission. 
Using the principle of detailed balance together with the equilibrium density $dn_\gamma/d\omega = ( \omega^2/\pi^2)[\exp(\omega/T)-1]^{-1}$ yields%
\begin{align}
    \Gamma^{(Q)}_{\rm abs}(\omega,T) = \frac{\pi^2}{\omega^2}e^{\omega/T} \frac{d\Gamma^{(Q)}_\text{brem}}{d\omega dV}. 
\end{align}
Of course, this relation is not restricted to quadrupole transitions, but holds generally.
In ordinary units, and using the emissivity $j^{(Q)}_\nu(T)$ introduced above, the relation reads $\Gamma^{(Q)}_{\rm abs}(\nu,T) = c^3/(4\pi \hbar \nu^3) j^{(Q)}_\nu(T)$. 

Finally, the absorption opacity is defined as the \textit{net} rate of absorption of photons of energy $\omega$ minus the rate per initial photon of photons emitted at the same energy by stimulated emission.  Hence, $c\rho\kappa^{(Q)}_{\text{ff}} = \Gamma^{(Q)}_{\rm abs} [1-\exp(-h\nu/k_{\rm B}T)]$ so that the coefficient of the frequency-dependent absorption opacity due to the quadrupole process reads, 
\begin{align}
   \kappa^{(Q)}_{\text{ff}} 
    \simeq 
    \kappa_0 \rho N_e^2 \nu_{\rm GHz}^{-3} T_8^{1/2}  \langle \gff \rangle  
    \left(1 - e^{-3.015\times 10^{-9}\nu_{\rm GHz}/T_8} \right)
    .
\end{align}
with $\kappa_0 = 9.10\times 10^{-26}\, {\rm cm^5}$; $\rho$ is the mass density of the medium in $\rm g/cm^3$ and $N_e = n_e/\rho$ is the number of electrons per gram; $1/(\rho \kappa_{\rm abs})$ has dimension of length. If one wishes to compute the quadrupole opacity, say from electrons and ions $I$ of charge $Z$, one replaces $N_e^2$ above by $Z^2 N_e N_{I}/(2\sqrt{2})$ and uses the tabulation of the Gaunt factor $ \langle \gff \rangle$  for attractive interactions. In the above replacement, a factor $1/(4\sqrt{2})$ comes from the modified reduced mass.

\subsection{Numerical Evaluation}
\label{sec:num_gaunt_th}

We evaluate the integral~\eqref{eq:ThermSigma} for the range
 $10^{-3} \le |\nu_p| \le 10^{3}$ and $10^{-11}\le \xi \le 10$ using the tabulations obtained in Sec.~\ref{sec:num_gaunt}.
 The range in $\nu_p$ covers a great range in temperatures, $T\gtrsim 10^{-5}\, \text{eV}$ for electrons and $T\gtrsim 0.5 \,\text{eV}$ for the mutual scattering of fully ionized helium.
 Concretely, we linearly interpolate $\gnonid$ in the interval $\log_{10} x=[-12-\epsilon,-\epsilon]$ and $\log_{10} |\nu_i|=[-3,4]$ on the logarithmic grid. For $\gint$, we switch to a denser grid of $0.002 \, \text{dex}$ for $\log_{10} x \geq -2$, in order to resolve the oscillations of the \lss{exchange} term visible in the right panel of Fig.~\ref{fig:g_ff} and interpolate in the interval $\log_{10} |\nu_i|=[-3,0.9]$. 
For $|\nu_i|<10^{-3}$ the Born result can be used as it is accurate to at least six significant digits.
Arguments $x<10^{-12}$ and $|\nu_i|> 10 ^4$ that are formally required in the integral~\eqref{eq:ThermSigma} are not available to us. For the tabulated range, we estimate for the relative error $|\Delta\langle\gff\rangle/\langle\gff\rangle|<10^{-4}$  %
 for $\log_{10} \xi \ge -11$  and  $\log_{10} |\nu_p| \leq 3$.
We do so by making the  antipodal choices of setting $d\sigma_{Q}/dx = 0$ and  $d\sigma_{Q}/dx = d\sigma_{Q}/dx|_{
\rm Born}$ outside the tabulated region. The good accuracy is owed to the fact the integrand $u \, \exp(-u)$ is peaked at $u=1$ so that function values beyond the boundaries of the table are unimportant. 
\lss{We note in passing that the error induced by the bi-linear interpolation of the tabulated values is negligible with respect to the target tolerance of the integration routine.}

The results for $\langle\gnonid\rangle$ are shown for repulsive ($Z_1 Z_2>0$) and attractive ($Z_1 Z_2<0$) interactions  in the left and right panel of Fig.~\ref{fig:gth_ff}, respectively.
For soft photons $\xi \ll 1$ the thermally averaged Gaunt factors for attractive and repulsive interactions coincide because in this limit $e^{-2\pi \nu_i} S_f/S_i \to 1$. For $\xi \gtrsim \nu_p^{-1}$, the Gaunt factor decreases for repulsive interactions.
For attractive interactions, the Gaunt factor increases for $\xi \gtrsim \nu_p^{-1}$ and reaches its maximum at $\xi=1$, above which the Boltzmann-suppression leads to a strong decline. 
Figure~\ref{fig:gth_int} shows the thermally averaged \lss{exchange} term $\langle\gint\rangle$. Note that $\langle\gint\rangle$ can have negative values, indicated by red line-segments, but $|\langle\gint\rangle|<\langle\gnonid\rangle$ always, such that the cross section is always positive.
The three thermally averaged Gaunt factors, $\langle\gnonid\rangle$ for repulsive and attractive interactions, as well as $\langle\gint\rangle$ are tabulated as a function of $|\nu_p|$ in the supplementary electronic tables with four significant digits of precision; an excerpt is shown in Tab.~\ref{tab:3}. In addition, the thermally averaged electron-electron Gaunt factor on the double logarithmic grid in temperature $T$ and emitted photon energy $\omega$ is provided in Tab.~\ref{tab:gthee}.

\begin{table*}[tp]
\centering
\setlength{\tabcolsep}{6.5pt}
\renewcommand{\arraystretch}{\tablesepvert}
\begin{tabular}{r | lllllllllll } 
\multicolumn{12}{c}{Thermally averaged gaunt factor $\langle\gnonid\rangle$ for $Z_1Z_2>0$ (top), $Z_1Z_2<0$ (middle), and $\langle\gint\rangle$ (bottom)} \\[3pt]
 \toprule
  \multicolumn{1}{c}{}&\multicolumn{11}{c}{$\log_{10}|\nu_p|$} \\
  \multicolumn{1}{c|}{$\log_{10}\xi$} &
   \multicolumn{1}{c}{\Minus2.0} & \multicolumn{1}{c}{\Minus1.5} & \multicolumn{1}{c}{\Minus1.0} & \multicolumn{1}{c}{\Minus0.5} & \multicolumn{1}{c}{0.0} & \multicolumn{1}{c}{0.5} & \multicolumn{1}{c}{1.0} & \multicolumn{1}{c}{1.5} & \multicolumn{1}{c}{2.0} & \multicolumn{1}{c}{2.5} & \multicolumn{1}{c}{3.0}   \\
 \midrule
\Minus11.0 & 9.21\Plus1 & 9.21\Plus1 & 9.21\Plus1 & 9.18\Plus1 & 9.03\Plus1 & 8.72\Plus1 & 8.36\Plus1 & 7.99\Plus1 & 7.62\Plus1 & 7.25\Plus1 & 6.89\Plus1 \\
\Minus10.0 & 8.48\Plus1 & 8.48\Plus1 & 8.48\Plus1 & 8.45\Plus1 & 8.30\Plus1 & 7.99\Plus1 & 7.63\Plus1 & 7.26\Plus1 & 6.89\Plus1 & 6.52\Plus1 & 6.15\Plus1 \\
\Minus9.0 & 7.74\Plus1 & 7.74\Plus1 & 7.74\Plus1 & 7.71\Plus1 & 7.57\Plus1 & 7.25\Plus1 & 6.89\Plus1 & 6.52\Plus1 & 6.15\Plus1 & 5.79\Plus1 & 5.42\Plus1 \\
\Minus8.0 & 7.01\Plus1 & 7.01\Plus1 & 7.00\Plus1 & 6.98\Plus1 & 6.83\Plus1 & 6.52\Plus1 & 6.15\Plus1 & 5.78\Plus1 & 5.42\Plus1 & 5.05\Plus1 & 4.68\Plus1 \\
\Minus7.0 & 6.27\Plus1 & 6.27\Plus1 & 6.27\Plus1 & 6.24\Plus1 & 6.09\Plus1 & 5.78\Plus1 & 5.42\Plus1 & 5.05\Plus1 & 4.68\Plus1 & 4.31\Plus1 & 3.94\Plus1 \\
\Minus6.0 & 5.53\Plus1 & 5.53\Plus1 & 5.53\Plus1 & 5.50\Plus1 & 5.36\Plus1 & 5.04\Plus1 & 4.68\Plus1 & 4.31\Plus1 & 3.94\Plus1 & 3.57\Plus1 & 3.20\Plus1 \\
\Minus5.0 & 4.80\Plus1 & 4.80\Plus1 & 4.79\Plus1 & 4.76\Plus1 & 4.62\Plus1 & 4.30\Plus1 & 3.94\Plus1 & 3.57\Plus1 & 3.20\Plus1 & 2.82\Plus1 & 2.43\Plus1 \\
\Minus4.0 & 4.06\Plus1 & 4.06\Plus1 & 4.05\Plus1 & 4.03\Plus1 & 3.88\Plus1 & 3.57\Plus1 & 3.20\Plus1 & 2.82\Plus1 & 2.43\Plus1 & 2.01\Plus1 & 1.53\Plus1 \\
\Minus3.0 & 3.32\Plus1 & 3.32\Plus1 & 3.32\Plus1 & 3.29\Plus1 & 3.14\Plus1 & 2.81\Plus1 & 2.43\Plus1 & 2.01\Plus1 & 1.53\Plus1 & 9.96\Plus0 & 4.78\Plus0 \\
\Minus2.0 & 2.57\Plus1 & 2.57\Plus1 & 2.56\Plus1 & 2.53\Plus1 & 2.36\Plus1 & 1.99\Plus1 & 1.52\Plus1 & 9.92\Plus0 & 4.76\Plus0 & 1.31\Plus0 & 1.38\Minus1 \\
\Minus1.0 & 1.75\Plus1 & 1.75\Plus1 & 1.72\Plus1 & 1.65\Plus1 & 1.40\Plus1 & 9.43\Plus0 & 4.55\Plus0 & 1.25\Plus0 & 1.32\Minus1 & 2.79\Minus3 & 4.41\Minus6 \\
0.0 & 6.02\Plus0 & 5.90\Plus0 & 5.52\Plus0 & 4.53\Plus0 & 2.61\Plus0 & 7.69\Minus1 & 8.22\Minus2 & 1.76\Minus3 & 2.79\Minus6 & 6.82\Minus11 & 2.16\Minus18 \\
1.0 & 1.21\Minus3 & 1.15\Minus3 & 9.74\Minus4 & 5.89\Minus4 & 1.44\Minus4 & 5.63\Minus6 & 1.34\Minus8 & 4.38\Minus13 & 1.69\Minus20 & 1.57\Minus32 & 5.82\Minus52 \\

 \midrule

\Minus11.0 & 9.21\Plus1 & 9.21\Plus1 & 9.21\Plus1 & 9.18\Plus1 & 9.03\Plus1 & 8.72\Plus1 & 8.36\Plus1 & 7.99\Plus1 & 7.62\Plus1 & 7.25\Plus1 & 6.89\Plus1 \\
\Minus10.0 & 8.48\Plus1 & 8.48\Plus1 & 8.48\Plus1 & 8.45\Plus1 & 8.30\Plus1 & 7.99\Plus1 & 7.63\Plus1 & 7.26\Plus1 & 6.89\Plus1 & 6.52\Plus1 & 6.15\Plus1 \\
\Minus9.0 & 7.74\Plus1 & 7.74\Plus1 & 7.74\Plus1 & 7.71\Plus1 & 7.57\Plus1 & 7.25\Plus1 & 6.89\Plus1 & 6.52\Plus1 & 6.15\Plus1 & 5.79\Plus1 & 5.42\Plus1 \\
\Minus8.0 & 7.01\Plus1 & 7.01\Plus1 & 7.00\Plus1 & 6.98\Plus1 & 6.83\Plus1 & 6.52\Plus1 & 6.15\Plus1 & 5.78\Plus1 & 5.42\Plus1 & 5.05\Plus1 & 4.68\Plus1 \\
\Minus7.0 & 6.27\Plus1 & 6.27\Plus1 & 6.27\Plus1 & 6.24\Plus1 & 6.09\Plus1 & 5.78\Plus1 & 5.42\Plus1 & 5.05\Plus1 & 4.68\Plus1 & 4.31\Plus1 & 3.94\Plus1 \\
\Minus6.0 & 5.53\Plus1 & 5.53\Plus1 & 5.53\Plus1 & 5.50\Plus1 & 5.36\Plus1 & 5.04\Plus1 & 4.68\Plus1 & 4.31\Plus1 & 3.94\Plus1 & 3.58\Plus1 & 3.21\Plus1 \\
\Minus5.0 & 4.80\Plus1 & 4.80\Plus1 & 4.79\Plus1 & 4.76\Plus1 & 4.62\Plus1 & 4.31\Plus1 & 3.94\Plus1 & 3.58\Plus1 & 3.21\Plus1 & 2.85\Plus1 & 2.51\Plus1 \\
\Minus4.0 & 4.06\Plus1 & 4.06\Plus1 & 4.06\Plus1 & 4.03\Plus1 & 3.88\Plus1 & 3.57\Plus1 & 3.21\Plus1 & 2.85\Plus1 & 2.51\Plus1 & 2.19\Plus1 & 1.93\Plus1 \\
\Minus3.0 & 3.32\Plus1 & 3.32\Plus1 & 3.32\Plus1 & 3.29\Plus1 & 3.15\Plus1 & 2.84\Plus1 & 2.50\Plus1 & 2.19\Plus1 & 1.93\Plus1 & 1.79\Plus1 & 1.87\Plus1 \\
\Minus2.0 & 2.57\Plus1 & 2.57\Plus1 & 2.57\Plus1 & 2.56\Plus1 & 2.43\Plus1 & 2.17\Plus1 & 1.92\Plus1 & 1.78\Plus1 & 1.86\Plus1 & 2.35\Plus1 & 3.64\Plus1 \\
\Minus1.0 & 1.76\Plus1 & 1.76\Plus1 & 1.78\Plus1 & 1.80\Plus1 & 1.77\Plus1 & 1.69\Plus1 & 1.76\Plus1 & 2.20\Plus1 & 3.37\Plus1 & 6.04\Plus1 & 1.19\Plus2 \\
0.0 & 6.14\Plus0 & 6.27\Plus0 & 6.67\Plus0 & 7.69\Plus0 & 9.05\Plus0 & 1.09\Plus1 & 1.55\Plus1 & 2.63\Plus1 & 5.00\Plus1 & 1.01\Plus2 & 2.12\Plus2 \\
1.0 & 1.27\Minus3 & 1.34\Minus3 & 1.58\Minus3 & 2.35\Minus3 & 3.95\Minus3 & 5.53\Minus3 & 8.31\Minus3 & 1.45\Minus2 & 2.81\Minus2 & 5.75\Minus2 & 1.21\Minus1 \\

 \midrule

\Minus11.0 & 1.60\Plus0 & 1.60\Plus0 & 1.57\Plus0 & 1.39\Plus0 & 6.63\Minus1 & 3.74\Minus2 & \multicolumn{1}{c}{\dots} & \multicolumn{1}{c}{\dots} & \multicolumn{1}{c}{\dots} & \multicolumn{1}{c}{\dots} & \multicolumn{1}{c}{\dots} \\
\Minus10.0 & 1.60\Plus0 & 1.60\Plus0 & 1.57\Plus0 & 1.39\Plus0 & 6.64\Minus1 & 3.77\Minus2 & \multicolumn{1}{c}{\dots} & \multicolumn{1}{c}{\dots} & \multicolumn{1}{c}{\dots} & \multicolumn{1}{c}{\dots} & \multicolumn{1}{c}{\dots} \\
\Minus9.0 & 1.60\Plus0 & 1.60\Plus0 & 1.57\Plus0 & 1.39\Plus0 & 6.64\Minus1 & 3.77\Minus2 & \multicolumn{1}{c}{\dots} & \multicolumn{1}{c}{\dots} & \multicolumn{1}{c}{\dots} & \multicolumn{1}{c}{\dots} & \multicolumn{1}{c}{\dots} \\
\Minus8.0 & 1.60\Plus0 & 1.60\Plus0 & 1.57\Plus0 & 1.39\Plus0 & 6.64\Minus1 & 3.77\Minus2 & \multicolumn{1}{c}{\dots} & \multicolumn{1}{c}{\dots} & \multicolumn{1}{c}{\dots} & \multicolumn{1}{c}{\dots} & \multicolumn{1}{c}{\dots} \\
\Minus7.0 & 1.60\Plus0 & 1.60\Plus0 & 1.57\Plus0 & 1.39\Plus0 & 6.64\Minus1 & 3.77\Minus2 & \multicolumn{1}{c}{\dots} & \multicolumn{1}{c}{\dots} & \multicolumn{1}{c}{\dots} & \multicolumn{1}{c}{\dots} & \multicolumn{1}{c}{\dots} \\
\Minus6.0 & 1.60\Plus0 & 1.60\Plus0 & 1.57\Plus0 & 1.39\Plus0 & 6.64\Minus1 & 3.77\Minus2 & \multicolumn{1}{c}{\dots} & \multicolumn{1}{c}{\dots} & \multicolumn{1}{c}{\dots} & \multicolumn{1}{c}{\dots} & \multicolumn{1}{c}{\dots} \\
\Minus5.0 & 1.60\Plus0 & 1.60\Plus0 & 1.57\Plus0 & 1.39\Plus0 & 6.64\Minus1 & 3.76\Minus2 & \multicolumn{1}{c}{\dots} & \multicolumn{1}{c}{\dots} & \multicolumn{1}{c}{\dots} & \multicolumn{1}{c}{\dots} & \multicolumn{1}{c}{\dots} \\
\Minus4.0 & 1.60\Plus0 & 1.60\Plus0 & 1.57\Plus0 & 1.39\Plus0 & 6.64\Minus1 & 3.76\Minus2 & \multicolumn{1}{c}{\dots} & \multicolumn{1}{c}{\dots} & \multicolumn{1}{c}{\dots} & \multicolumn{1}{c}{\dots} & \multicolumn{1}{c}{\dots} \\
\Minus3.0 & 1.60\Plus0 & 1.60\Plus0 & 1.57\Plus0 & 1.39\Plus0 & 6.63\Minus1 & 3.76\Minus2 & \multicolumn{1}{c}{\dots} & \multicolumn{1}{c}{\dots} & \multicolumn{1}{c}{\dots} & \multicolumn{1}{c}{\dots} & \multicolumn{1}{c}{\dots} \\
\Minus2.0 & 1.59\Plus0 & 1.59\Plus0 & 1.56\Plus0 & 1.38\Plus0 & 6.55\Minus1 & 3.71\Minus2 & \multicolumn{1}{c}{\dots} & \multicolumn{1}{c}{\dots} & \multicolumn{1}{c}{\dots} & \multicolumn{1}{c}{\dots} & \multicolumn{1}{c}{\dots} \\
\Minus1.0 & 1.57\Plus0 & 1.55\Plus0 & 1.51\Plus0 & 1.27\Plus0 & 5.67\Minus1 & 3.15\Minus2 & \multicolumn{1}{c}{\dots} & \multicolumn{1}{c}{\dots} & \multicolumn{1}{c}{\dots} & \multicolumn{1}{c}{\dots} & \multicolumn{1}{c}{\dots} \\
0.0 & 1.30\Plus0 & 1.25\Plus0 & 1.11\Plus0 & 7.84\Minus1 & 2.27\Minus1 & 5.08\Minus3 & \multicolumn{1}{c}{\dots} & \multicolumn{1}{c}{\dots} & \multicolumn{1}{c}{\dots} & \multicolumn{1}{c}{\dots} & \multicolumn{1}{c}{\dots} \\
1.0 & 8.08\Minus4 & 7.59\Minus4 & 6.25\Minus4 & 3.48\Minus4 & 6.79\Minus5 & 9.92\Minus7 & \multicolumn{1}{c}{\dots} & \multicolumn{1}{c}{\dots} & \multicolumn{1}{c}{\dots} & \multicolumn{1}{c}{\dots} & \multicolumn{1}{c}{\dots} \\

 \bottomrule
\end{tabular}
\tablecomments{\TableComment}
\caption{Thermally averaged free-free Gaunt factor for non-identical particles $\langle\gnonid\rangle$ for repulsive interactions $Z_1 Z_2 >0$ (top), attractive interactions  $Z_1 Z_2 <0$ (middle), and $\langle \gint\rangle $ (bottom) 
in the range $10^{-3}\le|\nu_p|\le 10^3$ in increments of 0.1~dex and $10^{-11}\le\xi\le 10$ in increments of 0.05~dex. For $|\nu_p|\ge 10$, the values of $\langle\gint\rangle$ are beyond the precision of the table, indicated by the dots, and set to zero in the electronic version; the notation follows Tab.~\ref{tab:1}.
\label{tab:3}}
\end{table*}

\begin{table*}[tb]
\centering
\setlength{\tabcolsep}{4.2pt}
\renewcommand{\arraystretch}{\tablesepvert}
\begin{tabular}{r | lllllllllll } 
\multicolumn{12}{c}{Thermally averaged electron-electron gaunt factor $\big\langle\gff\big\rangle$} \\[3pt]
 \toprule
  \multicolumn{1}{c|}{}&\multicolumn{11}{c}{$\log_{10}(T/{\rm K})$} \\
  \multicolumn{1}{c|}{$\log_{10}(\omega/{\rm eV})$} &
  \multicolumn{1}{c}{\Minus1.0} & \multicolumn{1}{c}{0.0} & \multicolumn{1}{c}{1.0} &\multicolumn{1}{c}{2.0} &\multicolumn{1}{c}{3.0} &\multicolumn{1}{c}{4.0} &  \multicolumn{1}{c}{5.0} & \multicolumn{1}{c}{6.0} & \multicolumn{1}{c}{7.0} & \multicolumn{1}{c}{8.0} & \multicolumn{1}{c}{9.0} \\
 \midrule
\Minus6.0 & 1.56\Minus8 & 3.73\Minus1 & 1.20\Plus1 & 2.58\Plus1 & 3.71\Plus1 & 4.81\Plus1 & 5.89\Plus1 & 6.82\Plus1 & 7.61\Plus1 & 8.35\Plus1 & 9.09\Plus1 \\
\Minus5.0 & 3.34\Minus24 & 8.20\Minus5 & 2.31\Plus0 & 1.71\Plus1 & 2.96\Plus1 & 4.08\Plus1 & 5.15\Plus1 & 6.09\Plus1 & 6.87\Plus1 & 7.62\Plus1 & 8.35\Plus1 \\
\Minus4.0 & 7.23\Minus66 & 1.01\Minus14 & 1.55\Minus2 & 6.56\Plus0 & 2.17\Plus1 & 3.34\Plus1 & 4.42\Plus1 & 5.35\Plus1 & 6.13\Plus1 & 6.88\Plus1 & 7.62\Plus1 \\
\Minus3.0 & 1.12\Minus213 & 5.35\Minus42 & 9.26\Minus9 & 3.54\Minus1 & 1.19\Plus1 & 2.57\Plus1 & 3.68\Plus1 & 4.61\Plus1 & 5.40\Plus1 & 6.14\Plus1 & 6.88\Plus1 \\
\Minus2.0 & \multicolumn{1}{c}{---} & 3.70\Minus128 & 6.24\Minus27 & 4.82\Minus5 & 2.19\Plus0 & 1.70\Plus1 & 2.93\Plus1 & 3.87\Plus1 & 4.66\Plus1 & 5.40\Plus1 & 6.14\Plus1 \\
\Minus1.0 & \multicolumn{1}{c}{---} & \multicolumn{1}{c}{---} & 2.12\Minus94 & 2.23\Minus17 & 9.00\Minus3 & 6.21\Plus0 & 2.13\Plus1 & 3.13\Plus1 & 3.92\Plus1 & 4.67\Plus1 & 5.41\Plus1 \\
0.0 & \multicolumn{1}{c}{---} & \multicolumn{1}{c}{---} & \multicolumn{1}{c}{---} & 1.21\Minus71 & 2.12\Minus11 & 2.03\Minus1 & 1.11\Plus1 & 2.36\Plus1 & 3.18\Plus1 & 3.93\Plus1 & 4.67\Plus1 \\
1.0 & \multicolumn{1}{c}{---} & \multicolumn{1}{c}{---} & \multicolumn{1}{c}{---} & \multicolumn{1}{c}{---} & 2.38\Minus60 & 7.33\Minus8 & 1.20\Plus0 & 1.44\Plus1 & 2.43\Plus1 & 3.19\Plus1 & 3.93\Plus1 \\
2.0 & \multicolumn{1}{c}{---} & \multicolumn{1}{c}{---} & \multicolumn{1}{c}{---} & \multicolumn{1}{c}{---} & \multicolumn{1}{c}{---} & 1.49\Minus53 & 6.03\Minus6 & 2.77\Plus0 & 1.56\Plus1 & 2.44\Plus1 & 3.19\Plus1 \\
3.0 & \multicolumn{1}{c}{---} & \multicolumn{1}{c}{---} & \multicolumn{1}{c}{---} & \multicolumn{1}{c}{---} & \multicolumn{1}{c}{---} & \multicolumn{1}{c}{---} & 8.75\Minus51 & 5.19\Minus5 & 3.89\Plus0 & 1.60\Plus1 & 2.44\Plus1 \\
4.0 & \multicolumn{1}{c}{---} & \multicolumn{1}{c}{---} & \multicolumn{1}{c}{---} & \multicolumn{1}{c}{---} & \multicolumn{1}{c}{---} & \multicolumn{1}{c}{---} & \multicolumn{1}{c}{---} & 4.63\Minus50 & 1.15\Minus4 & 4.36\Plus0 & 1.61\Plus1 \\
5.0 & \multicolumn{1}{c}{---} & \multicolumn{1}{c}{---} & \multicolumn{1}{c}{---} & \multicolumn{1}{c}{---} & \multicolumn{1}{c}{---} & \multicolumn{1}{c}{---} & \multicolumn{1}{c}{---} & \multicolumn{1}{c}{---} & 7.07\Minus50 & 1.47\Minus4 & 4.52\Plus0 \\
 \bottomrule
\end{tabular}
\tablecomments{\TableComment}
\caption{Thermally averaged free-free gaunt factor $\big\langle\gff\big\rangle$ for electron electron bremsstrahlung in the range $0.1\,\text{K}\le T \le 10^9 \, \text{K}$ and $1\,\mu\text{eV}<\omega<10 \, \text{keV}$  on a logarithmic grid in increments of 0.1~dex.  The dash indicates that the values are too small to be represented as double precision numbers and are set to zero in the electronic version of the table; the notation follows Tab.~\ref{tab:1}.
\label{tab:gthee}}
\end{table*}

\subsection{Comparison with approximate expressions}

\begin{figure}[tb]
		\includegraphics[width=\columnwidth]{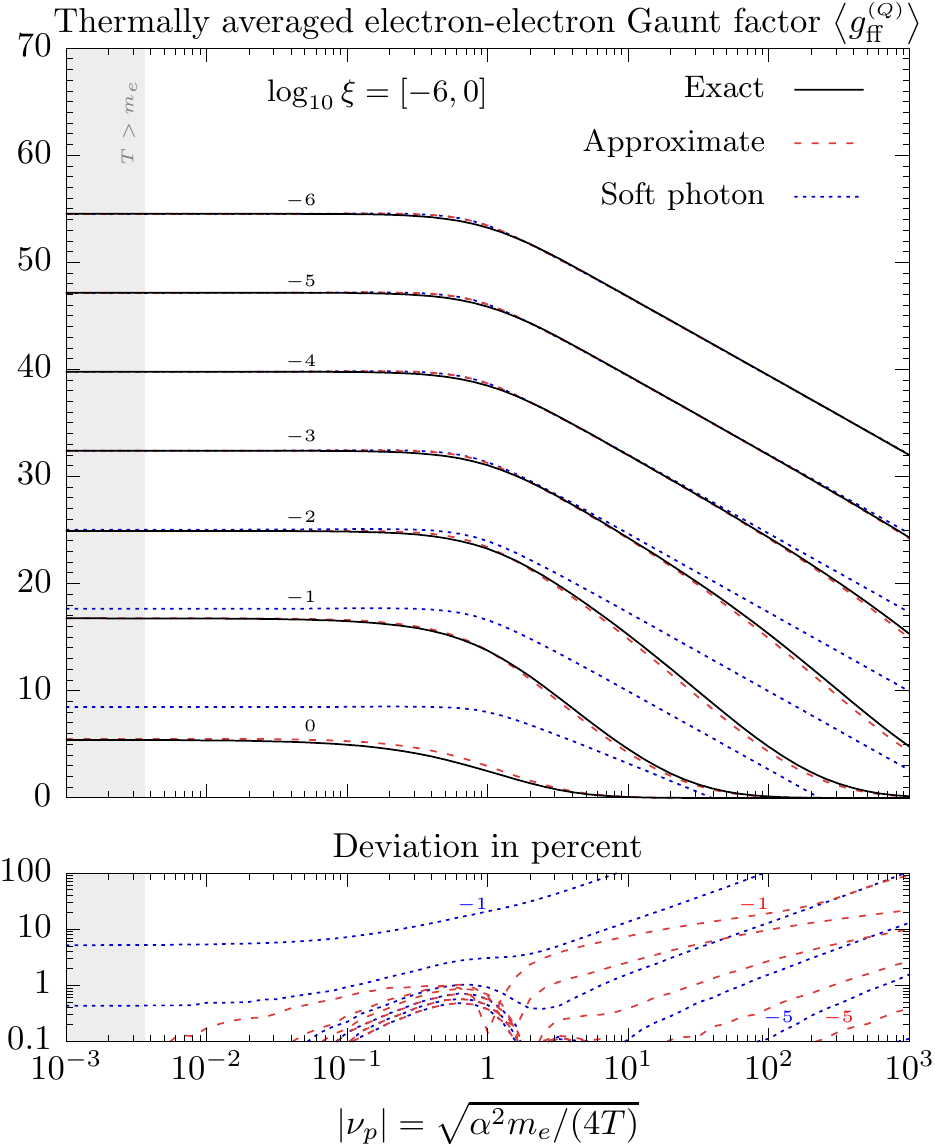}
		\caption{
		\emph{Top:} Thermally averaged electron-electron Gaunt factor over a wide range of parameters $\xi$ and $\nu_p$. The exact result (solid black lines) is compared to the one obtained from the soft photon limit (dotted blue) and the approximate formula~\eqref{eqn:quad_approximate} (dashed red). Inside the gray shaded region $T>m_e$ for electrons, which therefore lies outside the range of validity in the non-relativistic treatment.
		\lss{
		\emph{Bottom:} absolute percent deviation of the soft photon limit and the approximate formula with respect to the exact result for $\log_{10}\xi = [-5,-1]$. The labels of $\log_{10}\xi$ on the lines are to guide the eye. The deviation for $\log_{10}\xi = -6$ is below 1\%.
		}
		\label{fig:gth_soft_ff}}
\end{figure}
We are now in a position to compare the thermal average of the exact result with the one using the approximate expression~\eqref{eqn:quad_approximate} which is numerically easily obtained. For concreteness, we choose electron-electron scattering for which  $\langle\gff\rangle = \langle\gnonid\rangle - \langle\gint\rangle/2$.  Figure~\ref{fig:gth_soft_ff} shows the exact and approximate results  for multiple values of $\xi$ as solid black and dashed red lines, respectively. In addition, the dotted blue lines show the Gaunt factor derived from the soft photon approximation; see App.~\ref{app:soft_photon} for details.

One observes that the error of the soft photon approximation is \lss{smaller than 0.1\%} for $\xi=10^{-6}$ across the range $10^{-3}<|\nu_p|<10^3$. For $\xi\gtrsim 10^{-4}$, the soft photon approximation visibly deviates from the exact results for large $|\nu_p|$, but the error is still negligible for small $|\nu_p|$. The reason  is that Eq.~\eqref{eqn:quad_emission_weinberg} is valid for $x\ll \min\{1,|\nu_i|^{-1}\}$ (see \cite{Weinberg:2019mai} and \cite{Pradler:2020znn} for an in-depth discussion), which translates into the condition $\xi \ll \min\{1,|\nu_p|^{-1}\}$ for the thermally averaged quantities.
If this condition is not satisfied, the approximation breaks down and for large $|\nu_p|$ the soft Gaunt factor even yields unphysical negative values. Those problems are evaded by using  formula~\eqref{eqn:quad_approximate}. Not only does it extend the range of validity to larger $\xi$ in the Born limit, but also serves as an excellent approximation in the semi-classical limit and stays positive definite across the whole parameter range, as shown by the dashed red lines.

\section{Cooling function}
\label{sec:tot_gaunt}

\begin{figure}[tb]
		\includegraphics[width=\columnwidth]{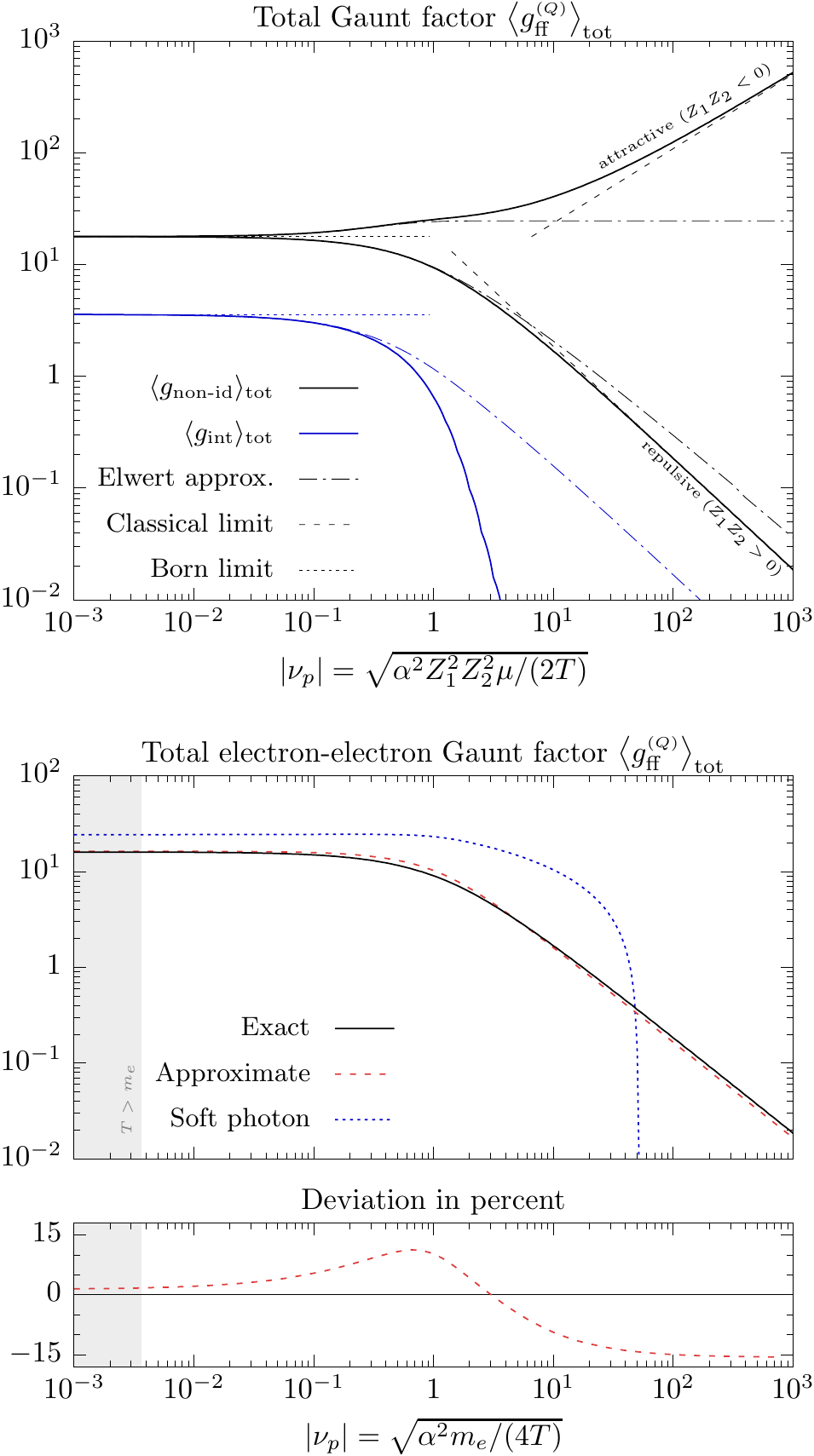}
		\caption{\emph{Top:} total free-free Gaunt factor for non-identical particles (black), and for the \lss{exchange} term (blue) over a wide $\nu_p$ range. The exact results (solid lines) are compared to the Elwert approximation (dash-dotted) as well as the Born (dotted) and classical (dashed) approximations. 
		\emph{Center:} total free-free Gaunt factor for identical spin-1/2 particles, \eg, electrons with $Z_{1,2}=1$ and $\mu = m_e/2$. The exact result (solid black line) is compared to one obtained from the soft photon limit (dotted blue) and from the approximate formula~\eqref{eqn:quad_approximate} (dashed red). Inside the gray shaded region $T>m_e$ for electrons, which  therefore lies outside the range of validity in the non-relativistic treatment.
		\lss{
		\emph{Bottom:} percent deviation of the approximate formula with respect to the exact result.
		}
		\label{fig:gtot_ff}}
\end{figure}

\subsection{Definition}
\label{sec:def_gaunt_tot}

In this section we provide the contribution of quadrupole radiation to the cooling function $\Lambda_{\rm brem}$,~\ie, the energy lost per volume and per time through quadru\-polar radiation,
\begin{subequations}
\begin{align}
&\Lambda_{\rm brem}^{(Q)} (\nu_p)
= \int_0^\infty d\omega \, \omega\,\frac{d\Gamma^{(Q)}_{\rm brem}}{dV d\omega} 
\\
   & \qquad= \sqrt{\frac{8}{\pi}}\frac{n_1 n_2}{(1+\delta_{12})}
    \frac{\alpha^3 Z_1^2 Z_2^2 A_Q^2}{\mu^{5/2}} 
    \,
    T^{3/2} \langle  \gff  \rangle_\text{tot} .
\end{align}
\end{subequations}
The second relation defines the total free-free Gaunt factor in accordance with literature~\citep{1961ApJS....6..167K, vanHoof:2014bha,Chluba:2019ser},
\begin{align} \label{eq:def_gtot}
    \langle  \gff  \rangle_\text{tot} (\nu_p) = 
    \int_0^\infty d\xi \,\langle  \gff  \rangle (\xi, \nu_p) .
\end{align}

In the Born and classical limits, the luminosity of quadrupole bremsstrahlung can be calculated analytically by executing the double integral in $\xi$ and $u$.
The resulting expressions are independent of $\nu_p$ and given by 
\begin{subequations}\label{eq:born_tot}
\begin{align}\label{eq:born_tot_nonid}
   \left\langle  \gnonid  \right\rangle_\text{tot} \Big|_{\text{Born}}&=\frac{160}{9} ,
   \\
   \left\langle 
   \gint  \right\rangle_\text{tot} \Big|_{\text{Born}}&= \frac{20}{9}\left(3\pi^2-28\right).
\end{align}
\end{subequations}
In the semi-classical limit, \lss{exchange} effects are not present and the Gaunt factor for a repulsive potential is \citep{landau1975classical}
\begin{align}\label{eq:class_tot}
   \langle  \gff  \rangle_\text{tot} \Big|_{\text{classical}}&=\frac{10 }{3}\frac{\pi^{3/2}}{|\nu_p|} .
\end{align}
It follows from Eqs.~\eqref{eq:born_tot} and~\eqref{eq:class_tot} that the cooling function scales with temperature as $\Lambda_{\rm brem}^{(Q)} \sim T^{3/2}$ in the Born limit and as $\Lambda_{\rm brem}^{(Q)} \sim T^{2}$ in the semi-classical limit. 
For attractive interactions, the semi-classical limit of the total Gaunt factor is \citep{Pradler:2020znn}
\begin{align}\label{eq:class_tot_att}
   \langle  \gff  \rangle_\text{tot} \Big|_{\text{classical}}&=
	\frac{16}{5}\frac{ 2^{1/3} \pi^{3/2} }{3^{1/6}} \frac{\Gamma(8/3)}{\Gamma(1/6)} \nu_p^{2/3}
	 .
\end{align}
Hence, the cooling function scales with temperature as $\Lambda_{\rm brem}^{(Q)} \sim T^{7/6}$ in the classical regime.

In absolute numbers, and specializing to the case of electron-electron bremsstrahlung the contribution to the cooling function reads
\begin{align}
   \Lambda_{\rm brem}^{(Q)}
     = 
    %  3.51 \times 10^{-26}\,
    %  \, T_8^{3/2}
    %  n_e^2 \, \langle 
    %  \gff \rangle_{\rm tot} \: {\rm erg\ cm^3\ s^{-1}} .
     3.51 \times 10^{-26}\,
     \, T_8^{3/2}
    \lss{\left(\frac{n_e}{\rm cm^3}\right)^2} \, \langle 
     \gff \rangle_{\rm tot} \:
    \lss{ \frac{\rm erg}{\rm cm^3\, s}}.
     %{\rm erg\ cm^{-3}\ s^{-1}} .
\end{align}
where  $ \langle 
     \gff \rangle_{\rm tot}$ is evaluated at $|\nu_p|=0.028T_8^{-1/2}$. If one wishes to compute the quadrupole bremsstrahlung contribution for electron-ion scattering one replaces $n_e^2$ by $Z^2 n_e n_I/(2\sqrt{2})$ and uses the tabulation of  $\langle 
     \gff \rangle_{\rm tot}$ for attractive interactions. 
     Of course, either of the latter processes only contribute a small fraction to the free-free cooling function which is dominated by electron dipole-bremsstrahlung; for electron-ion scattering, their relative ratio reads,
\begin{align}
\label{coolnum}
    \left. \frac{ \Lambda_{\rm brem}^{(Q)} }{ \Lambda_{\rm brem}^{(D)} }\right|_{\text{e-ion}} = 
    3.5
    \times 10^{-3} T_8 \,
     \frac{\langle \gff \rangle_{\rm tot}}{ \langle g_{\rm ff}^{(D)} \rangle_{\rm tot }} .
\end{align}
with $\langle g_{\rm ff}^{(D)} \rangle_{\rm tot }$ defined as usual~\citep{1961ApJS....6..167K}. 
On the account that the (temperature-dependent) ratio of Gaunt factors can be $\sim 10-100$, quadrupole electron-ion brems\-strahlung makes a few percent correction in a plasma of 10~keV temperature ($\nu_p\lesssim 1$), and accordingly less in colder environments; similar conclusions hold when electron-electron quadrupole bremsstrahlung is compared to the electron-ion dipole process for which an additional factor ${2\sqrt{2}\, n_e}/(Z^2 n_I)$ appears in~\eqref{coolnum}.

\subsection{Numerical Evaluation}
\label{sec:num_gaunt_tot}

\begin{table}[tb]
\centering
\setlength{\tabcolsep}{5pt}
\renewcommand{\arraystretch}{\tablesepvert}
\begin{tabular}{r | r |r r | r} 
\multicolumn{5}{c}{Total Gaunt factor} \\[3pt]
 \toprule
 \multicolumn{1}{c|}{} &
  \multicolumn{1}{c|}{$e^- e^-$} &
 \multicolumn{2}{c|}{$Z_1 Z_2 >0$} &
  \multicolumn{1}{c}{$Z_1 Z_2 <0$} \\
  \multicolumn{1}{c|}{$\log_{10}|\nu_p|$} &
  \multicolumn{1}{c|}{$\left\langle  g_\text{ff}  \right\rangle_\text{tot}$} &
  \multicolumn{1}{c}{$\left\langle  \gnonid  \right\rangle_\text{tot}$} &
  \multicolumn{1}{c|}{$\left\langle  \gint  \right\rangle_\text{tot}$} &
  \multicolumn{1}{c}{$\left\langle  \gnonid  \right\rangle_\text{tot}$} \\
  \midrule
\Minus3.0 & \dots & 17.78 & 3.58 & 17.81 \\
\Minus2.5 & \dots & 17.75 & 3.56 & 17.84 \\
\Minus2.0 & 15.89 & 17.65 & 3.52 & 17.94 \\
\Minus1.5 & 15.66 & 17.35 & 3.38 & 18.25 \\
\Minus1.0 & 14.94 & 16.45 & 3.01 & 19.24 \\
\Minus0.5 & 12.99 & 14.05 & 2.12 & 21.77 \\
0.0 & 9.05 & 9.37 & 0.65 & 25.25 \\
0.5 & 4.44 & 4.45 & 0.02 & 29.47 \\
1.0 & 1.68 & 1.68 & \dots & 40.39 \\
1.5 & 0.57 & 0.57 & \dots & 66.48 \\
2.0 & 0.18 & 0.18 & \dots & 124.3\hphantom{0} \\
2.5 & 0.06 & 0.06 & \dots & 250.1\hphantom{0} \\
3.0 & 0.02 & 0.02 & \dots & 521.5\hphantom{0} \\

 \bottomrule
\end{tabular}
\tablecomments{\TableComment}
\caption{Total free-free Gaunt factor for electron-electron scattering, as well as for general repulsive and attractive interactions in the range $10^{-3} \le |\nu_p| \le 10^3$ in increments of 0.005~dex. In the second column, values for which $T>m_e$ are set to zero since they are beyond the non-relativistic approximation. In the 4th column, $\langle\gint\rangle$ is set to zero for $|\nu_p|\ge 10$ as before; the notation follows Tab.~\ref{tab:1}.
\label{tab:gtot}}

\end{table}

We evaluate~\eqref{eq:def_gtot} by linearly interpolating the Tab.~\ref{tab:3} on the logarithmic grid and integrate over $
\xi$ across the tabulated range, $10^{-11}<\xi<10$. We estimate the error by changing the boundaries in the numerical integration in the Born limit, where the integral can be done analytically and find that the relative error is below $10^{-4}$ in the tabulated $\nu_p$ range.
\lss{As in Sec.~\ref{sec:num_gaunt_th}, the error induced by interpolating the tabulated values is negligible with respect to the target tolerance of the integration routine.}
Beyond the Born limit, the kinematic suppression of the Gaunt factor  increases for large $\xi$, rendering the $\xi>10$ contribution to the integral and its relative error even smaller. Outside the tabulated region $|\nu_p|<10^{-3}$ and $|\nu_p|>10^3$ the exact Gaunt factor agrees with the respective Born and classical limits to three digits of precision, respectively. In the latter cases, the asymptotic expressions \eqref{eq:born_tot} and \eqref{eq:class_tot} can then be used.

The results for $\langle\gnonid\rangle_\text{tot}$ and $\langle\gint\rangle_\text{tot}$ are shown in the top panel of Fig.~\ref{fig:gtot_ff} and are presented in the supplementary electronic table---a portion of which is shown in Tab.~\ref{tab:gtot}---with four significant digits of precision in the interval $\log_{10}|\nu_p|=[-3,3]$.
As can be seen, the contribution of the \lss{exchange} term is suppressed by a factor 5 (10)  for spin-0 (spin-1/2) particles in the Born limit. The suppression increases with increasing temperature, a property which is beyond the Elwert approximation.
In the bottom panel, the results obtained in the soft-photon approximation and from the approximate formula~\eqref{eqn:quad_approximate} are compared with the exact total Gaunt factor for electron-electron scattering. As can be seen, the soft photon approximation is inadequate for  all values of $\nu_p$, while the approximate formula provides a good approximation, and deviates by \lss{$+1\%,\ +11\%$, and $-16$\%} for $|\nu_p|=10^{-3},\ 1$, and $10^3$ from the exact result, respectively.
For convenience, the electron-electron Gaunt factor is also tabulated in Tab.~\ref{tab:gtot}.

\section{Conclusions}
\label{sec:conclusions}

In this work we provide highly accurate tabulations for the non-relativistic Gaunt factor of quadrupolar bremsstrahlung emission over a wide range of relative incoming velocity of the scattering particles and for arbitrary allowed photon energy. The calculations are based on the differential cross section that is exact in the Coulomb interaction of the colliding particle pair and was recently obtained in~\cite{Pradler:2020znn}. Thermally averaged Gaunt factors are provided for a Maxwellian non-relativistic plasma, as well as a tabulation of the total Gaunt factor which enters in the evaluation of the cooling function. For the latter, quadruopole radiation can make up to a percent-level correction in an intracluster gas. 

\paragraph*{Acknowledgements: }\ \  The authors are supported by the New Frontiers program of the Austrian Academy of Sciences. LS was in part supported by the Austrian Science Fund FWF under the Doctoral Program W1252-N27 Particles and Interactions.

\appendix

\section{Thermal Photon Emission}
\label{app:thermal}

In the energy differential bremsstrahlung cross section as calculated in~\cite{Pradler:2020znn}, the emitted photon energy is in the CM frame. When considering dipole emission from non-relativistic electrons on ions, the medium- and CM-frame essentially coincide. This is \textit{a priori} not the case for electron-electron bremsstrahlung. To account for the mismatch between CM and medium frame, and to obtain the production spectrum of photons, \ie\ the number of photons  produced per unit volume per unit time and per unit energy interval in the medium frame one computes,
\begin{widetext}
\begin{align}
\label{eq:prodspecfull}
\frac{d\Gamma_{\rm brem}}{dV d\omega} &= 
\frac{1}{\pi} \frac{n_{1} n_{2}}{1+\delta_{12}} \left( \frac{\mu}{T}\right)^{3/2} \sqrt{\frac{M}{T}} e^{M/T} \int_{v_i^\text{min}}^\infty dv_i\, v_i^3 \int_0^\infty d\omega_*
\frac{1}{\omega_*^2}
e^{- \frac{M}{T} \left( 1+ \frac{\mu v_i^2}{2M} \right) \frac{(\omega^2+\omega_*^2)}{2\omega \omega_*}}
\left[ \omega_* \frac{d\sigma(\omega_*,v_i)}{d\omega_*}\right]     ,
\end{align}
\end{widetext}
where $M= m_1 + m_2$ is the total mass of the
colliding particles species ``1'' and ``2'' with their respective number densities $n_{1,2}$, distributed with Maxwell-Boltzmann form with assumed common temperature $T$ and  $\omega_{(*)}$ is the emitted energy in the medium (CM) frame; $v_i^{\rm min}$ is the minimum required relative initial velocity to produce a quantum of energy $\omega$.
We obtain this expression as the non-relativistic limit of the general formula~(22) given in~\cite{1984ApJ...280..328D}. In the second integral, \lss{in the kinematically unsuppressed regime} the \lss{second} factor  $  \omega_* \frac{d\sigma(\omega_*,v)}{d\omega_*}$ is a relatively mild function of $\omega_*$ whereas the \lss{first} factor is a strongly peaked function at $\omega_* = \omega$ on the account that $T/M$ and $v$ are small parameters. We may hence pull out the \lss{second} factor, evaluate it at $\omega_*=\omega$ and integrate the \lss{first} factor to a modified Bessel function of the second kind of degree one. Upon expansion in the small parameters of the latter, we arrive at
      {
\medmuskip=0mu
\begin{align} \label{eq:prodspec}
  \frac{d\Gamma_{\rm brem}}{dV d\omega} &\simeq  \frac{n_{1} n_{2}}{1+\delta_{12}}\sqrt{\frac{2}{\pi}}  \left(\frac{\mu}{ T} \right)^{3 / 2} \int_{v_i^\text{min}}^\infty d v_i v_i^3 e^{-\frac{\mu v_i^{2}}{2T}}
   \frac{d\sigma\left(\omega, v_i \right)}{d \omega} \nonumber \\
   &\equiv \frac{n_{1} n_{2}}{1+\delta_{12}}  \frac{d\langle\sigma v\rangle}{d \omega}  . 
\end{align}
}%
We hence recover the result that for non-relativistic collisions the CM photon spectrum is a good approximation to the emission spectrum in the medium frame which is the quantity of interest; in this approximation,  $v_{i}^{\rm min} = \sqrt{2\omega/\mu}$. The second equality defines the thermally averaged cross section.  If one strives for greater accuracy, one may evaluate~\eqref{eq:prodspecfull} instead.

\section{Born cross sections}
\label{app:born_cs}
The Gaunt factor in this work is defined with respect to the Born cross section for non-identical particles~\eqref{eq:bornnonid}. 
For identical particles, such as for electron-electron scattering, with mass $m_{1,2}=m$ and charge $Z_{1,2}=Z$ the Born cross section is~\citep{Fedyushin:1952hg,1981PhRvA..23.2851G,1990ApJ...362..284G},
\begin{widetext}
{
\medmuskip=0mu
\thinmuskip=1mu
\thickmuskip=1mu
\nulldelimiterspace=1pt
\scriptspace=1pt
\begin{align}
    x \left.\frac{d\sigma_Q}{dx} \right|_{\substack{\text{Born}\\ \text{id}}  }
    &=
    \frac{8}{15} \frac{\alpha^{3} Z^6}{m^2} \sqrt{1-x}
		\Bigg\{
		10 
		+ \frac{3 (2-x)}{\sqrt{1-x} }\ln \left(\frac{1+ \sqrt{1-x}}{1- \sqrt{1-x}}\right) +
		\frac{(-1)^{2s}}{2s+1}
		\left[
		3+\frac{3x^2}{(2-x)^2}
    +\frac{7(2-x)^2x^2+3x^4}{2(2-x)^3\sqrt{1-x}}\ln \left(\frac{1+ \sqrt{1-x}}{1- \sqrt{1-x}}\right)
		\right]
		\Bigg\},
\end{align}
}%
\end{widetext}
where $s=0$ or $1/2$ is the spin of the scattering particles.
From Eq.~\eqref{eq:GauntQuadrupole} and the Born cross section for identical particles the \lss{exchange} Gaunt factor $\gint$ in the Born limit then reads, 
\begin{align} \label{eqn:ginterf_born}
    \gint  \Big|_{\text{Born}} = 
    \frac{ 
    3+\frac{3x^2}{(2-x)^2}
    +\frac{7(2-x)^2x^2+3x^4}{2(2-x)^3\sqrt{1-x}}\ln \left(\frac{1+ \sqrt{1-x}}{1- \sqrt{1-x}}\right)}
    {
    10 
		+ \frac{3 (2-x)}{\sqrt{1-x} }\ln \left(\frac{1+ \sqrt{1-x}}{1- \sqrt{1-x}}\right)
	}.
\end{align}
In the same limit $\gnonid$ as well as the prefactor $e^{-2\pi \nu_i}S_f/S_i$ in Eq.~\eqref{eq:GauntQuadrupole} approach unity.

\section{Soft Photon Limit}
\label{app:soft_photon}
If the emitted photons are soft, \ie~$x\ll 1$, a rather simple analytic formula for the differential cross section $d\sigma/dx$ has been found, that has the correct asymptotic form in the Born limit $|\nu_i|\ll 1$ and the semi-classical limit $|\nu_i|\ll 1$ \citep{Pradler:2020znn},
\begin{widetext}
\begin{align} 
    \left.x \frac{d \sigma_{Q}}{dx}\right|_{\text{soft}} &=
    \frac{8\alpha^{3} Z_1^2 Z_2^2 \mu^2}{15}
    \left(\frac{Z_1}{m_1^2} +\frac{Z_2}{m_2^2} \right)^2
    \Bigg\{
    13-\frac{6+3\delta_{12}}{\zeta^2+1}
    +6\ln \left(\frac{4}{x \zeta}\right) 
    + \frac{\pi^2 \nu_i^2}{\sinh^2{\pi \nu_i}}\left[\frac{6+3\delta_{12}}{\zeta^2+1}-\frac{6+3\delta_{12}}{2}+ 6\ln \zeta
    \right]
    \Bigg\}, \label{eqn:quad_emission_weinberg}
\end{align}
\end{widetext}
with $\zeta=|\nu_i| e^{\gamma+1/2}$ and $\delta_{12}=(0)1$ for (non-)identical spin-1/2 particles. The approximate formula~\eqref{eqn:quad_approximate} stated in the main text is an extension of this expression to also cover hard photon emission.

\section{Further tables}
\label{app:addtables}
In this final appendix, we provide the tabulation on a linear grid in~$x$ for the Gaunt factor for non-identical particles and for the \lss{exchange} term in~Tab.~\ref{tab:6}.

\begin{table*}[htb]
\centering
\setlength{\tabcolsep}{7.2pt}
\renewcommand{\arraystretch}{\tablesepvert}
\begin{tabular}{r | rrrrrrrrrrr } 
\multicolumn{12}{c}{Gaunt factors $\gnonid$ (top) and $ \gint$ (bottom)} \\[3pt]
 \toprule
  \multicolumn{1}{c|}{}&\multicolumn{11}{c}{$\log_{10}|\nu_i|$} \\
  \multicolumn{1}{c|}{$x$} &
  \multicolumn{1}{c}{\Minus1.0} & \multicolumn{1}{c}{\Minus0.5} & \multicolumn{1}{c}{0.0} & \multicolumn{1}{c}{0.5} & \multicolumn{1}{c}{1.0} & \multicolumn{1}{c}{1.5} & \multicolumn{1}{c}{2.0} & \multicolumn{1}{c}{2.5} & \multicolumn{1}{c}{3.0} & \multicolumn{1}{c}{3.5} & \multicolumn{1}{c}{4.0}  \\
 \midrule

0.01 & 9.99\Minus1 & 9.87\Minus1 & 9.21\Minus1 & 8.05\Minus1 & 7.06\Minus1 & 6.61\Minus1 & 7.11\Minus1 & 9.31\Minus1 & 1.48\Plus0 & 2.69\Plus0 & 5.32\Plus0 \\
0.09 & 9.99\Minus1 & 9.90\Minus1 & 9.44\Minus1 & 8.94\Minus1 & 9.58\Minus1 & 1.24\Plus0 & 1.96\Plus0 & 3.56\Plus0 & 7.02\Plus0 & 1.45\Plus1 & 3.06\Plus1 \\
0.17 & 1.00\Plus0 & 9.96\Minus1 & 9.77\Minus1 & 9.97\Minus1 & 1.20\Plus0 & 1.76\Plus0 & 3.04\Plus0 & 5.84\Plus0 & 1.19\Plus1 & 2.49\Plus1 & 5.31\Plus1 \\
0.25 & 1.00\Plus0 & 1.00\Plus0 & 1.01\Plus0 & 1.10\Plus0 & 1.43\Plus0 & 2.25\Plus0 & 4.08\Plus0 & 8.06\Plus0 & 1.67\Plus1 & 3.52\Plus1 & 7.51\Plus1 \\
0.33 & 1.00\Plus0 & 1.01\Plus0 & 1.04\Plus0 & 1.19\Plus0 & 1.65\Plus0 & 2.74\Plus0 & 5.12\Plus0 & 1.03\Plus1 & 2.14\Plus1 & 4.54\Plus1 & 9.71\Plus1 \\
0.41 & 1.00\Plus0 & 1.01\Plus0 & 1.07\Plus0 & 1.29\Plus0 & 1.88\Plus0 & 3.22\Plus0 & 6.17\Plus0 & 1.25\Plus1 & 2.63\Plus1 & 5.58\Plus1 & 1.20\Plus2 \\
0.49 & 1.00\Plus0 & 1.02\Plus0 & 1.10\Plus0 & 1.39\Plus0 & 2.10\Plus0 & 3.72\Plus0 & 7.23\Plus0 & 1.48\Plus1 & 3.12\Plus1 & 6.64\Plus1 & 1.42\Plus2 \\
0.57 & 1.00\Plus0 & 1.02\Plus0 & 1.13\Plus0 & 1.48\Plus0 & 2.33\Plus0 & 4.22\Plus0 & 8.31\Plus0 & 1.72\Plus1 & 3.62\Plus1 & 7.73\Plus1 & 1.66\Plus2 \\
0.65 & 1.00\Plus0 & 1.02\Plus0 & 1.16\Plus0 & 1.58\Plus0 & 2.56\Plus0 & 4.73\Plus0 & 9.43\Plus0 & 1.96\Plus1 & 4.14\Plus1 & 8.85\Plus1 & 1.90\Plus2 \\
0.73 & 1.00\Plus0 & 1.03\Plus0 & 1.19\Plus0 & 1.67\Plus0 & 2.79\Plus0 & 5.25\Plus0 & 1.06\Plus1 & 2.20\Plus1 & 4.68\Plus1 & 1.00\Plus2 & 2.15\Plus2 \\
0.81 & 1.00\Plus0 & 1.03\Plus0 & 1.22\Plus0 & 1.77\Plus0 & 3.03\Plus0 & 5.79\Plus0 & 1.17\Plus1 & 2.46\Plus1 & 5.23\Plus1 & 1.12\Plus2 & 2.40\Plus2 \\
0.89 & 1.00\Plus0 & 1.04\Plus0 & 1.25\Plus0 & 1.87\Plus0 & 3.28\Plus0 & 6.34\Plus0 & 1.30\Plus1 & 2.72\Plus1 & 5.80\Plus1 & 1.24\Plus2 & 2.67\Plus2 \\
0.97 & 1.00\Plus0 & 1.04\Plus0 & 1.28\Plus0 & 1.97\Plus0 & 3.53\Plus0 & 6.91\Plus0 & 1.42\Plus1 & 3.00\Plus1 & 6.39\Plus1 & 1.37\Plus2 & 2.94\Plus2 \\

 \midrule

0.01 & 6.43\Minus2 & 5.57\Minus2 & 1.78\Minus2 & 6.38\Minus5 & \multicolumn{1}{c}{\dots} & \multicolumn{1}{c}{\dots} & \multicolumn{1}{c}{\dots} & \multicolumn{1}{c}{\dots} & \multicolumn{1}{c}{\dots} & \multicolumn{1}{c}{\dots} & \multicolumn{1}{c}{\dots} \\
0.09 & 9.27\Minus2 & 7.96\Minus2 & 2.42\Minus2 & 7.42\Minus5 & \multicolumn{1}{c}{\dots} & \multicolumn{1}{c}{\dots} & \multicolumn{1}{c}{\dots} & \multicolumn{1}{c}{\dots} & \multicolumn{1}{c}{\dots} & \multicolumn{1}{c}{\dots} & \multicolumn{1}{c}{\dots} \\
0.17 & 1.11\Minus1 & 9.49\Minus2 & 1.94\Minus2 & 2.47\Minus5 & \multicolumn{1}{c}{\dots} & \multicolumn{1}{c}{\dots} & \multicolumn{1}{c}{\dots} & \multicolumn{1}{c}{\dots} & \multicolumn{1}{c}{\dots} & \multicolumn{1}{c}{\dots} & \multicolumn{1}{c}{\dots} \\
0.25 & 1.31\Minus1 & 1.12\Minus1 & 9.80\Minus3 &  \Minus1.43\Minus5 & \multicolumn{1}{c}{\dots} & \multicolumn{1}{c}{\dots} & \multicolumn{1}{c}{\dots} & \multicolumn{1}{c}{\dots} & \multicolumn{1}{c}{\dots} & \multicolumn{1}{c}{\dots} & \multicolumn{1}{c}{\dots} \\
0.33 & 1.54\Minus1 & 1.35\Minus1 & 1.94\Minus3 &  \Minus8.99\Minus4 & \multicolumn{1}{c}{\dots} & \multicolumn{1}{c}{\dots} & \multicolumn{1}{c}{\dots} & \multicolumn{1}{c}{\dots} & \multicolumn{1}{c}{\dots} & \multicolumn{1}{c}{\dots} & \multicolumn{1}{c}{\dots} \\
0.41 & 1.84\Minus1 & 1.64\Minus1 & 1.60\Minus3 &  \Minus6.25\Minus4 & \multicolumn{1}{c}{\dots} & \multicolumn{1}{c}{\dots} & \multicolumn{1}{c}{\dots} & \multicolumn{1}{c}{\dots} & \multicolumn{1}{c}{\dots} & \multicolumn{1}{c}{\dots} & \multicolumn{1}{c}{\dots} \\
0.49 & 2.22\Minus1 & 2.02\Minus1 & 1.41\Minus2 & 1.09\Minus3 & \multicolumn{1}{c}{\dots} & \multicolumn{1}{c}{\dots} & \multicolumn{1}{c}{\dots} & \multicolumn{1}{c}{\dots} & \multicolumn{1}{c}{\dots} & \multicolumn{1}{c}{\dots} & \multicolumn{1}{c}{\dots} \\
0.57 & 2.70\Minus1 & 2.52\Minus1 & 4.45\Minus2 & 1.74\Minus3 & \multicolumn{1}{c}{\dots} & \multicolumn{1}{c}{\dots} & \multicolumn{1}{c}{\dots} & \multicolumn{1}{c}{\dots} & \multicolumn{1}{c}{\dots} & \multicolumn{1}{c}{\dots} & \multicolumn{1}{c}{\dots} \\
0.65 & 3.33\Minus1 & 3.17\Minus1 & 9.76\Minus2 &  \Minus1.26\Minus3 & \multicolumn{1}{c}{\dots} & \multicolumn{1}{c}{\dots} & \multicolumn{1}{c}{\dots} & \multicolumn{1}{c}{\dots} & \multicolumn{1}{c}{\dots} & \multicolumn{1}{c}{\dots} & \multicolumn{1}{c}{\dots} \\
0.73 & 4.17\Minus1 & 4.03\Minus1 & 1.77\Minus1 &  \Minus7.62\Minus3 & \multicolumn{1}{c}{\dots} & \multicolumn{1}{c}{\dots} & \multicolumn{1}{c}{\dots} & \multicolumn{1}{c}{\dots} & \multicolumn{1}{c}{\dots} & \multicolumn{1}{c}{\dots} & \multicolumn{1}{c}{\dots} \\
0.81 & 5.29\Minus1 & 5.17\Minus1 & 2.86\Minus1 &  \Minus1.32\Minus2 & \multicolumn{1}{c}{\dots} & \multicolumn{1}{c}{\dots} & \multicolumn{1}{c}{\dots} & \multicolumn{1}{c}{\dots} & \multicolumn{1}{c}{\dots} & \multicolumn{1}{c}{\dots} & \multicolumn{1}{c}{\dots} \\
0.89 & 6.82\Minus1 & 6.68\Minus1 & 4.21\Minus1 &  \Minus1.15\Minus2 & \multicolumn{1}{c}{\dots} & \multicolumn{1}{c}{\dots} & \multicolumn{1}{c}{\dots} & \multicolumn{1}{c}{\dots} & \multicolumn{1}{c}{\dots} & \multicolumn{1}{c}{\dots} & \multicolumn{1}{c}{\dots} \\
0.97 & 8.95\Minus1 & 8.68\Minus1 & 5.57\Minus1 & 6.75\Minus4 & \multicolumn{1}{c}{\dots} & \multicolumn{1}{c}{\dots} & \multicolumn{1}{c}{\dots} & \multicolumn{1}{c}{\dots} & \multicolumn{1}{c}{\dots} & \multicolumn{1}{c}{\dots} & \multicolumn{1}{c}{\dots} \\

 \bottomrule
\end{tabular}
\tablecomments{\TableComment}
\caption{
Free-free Gaunt factor for non-identical particles $\gnonid$ (top) and the \lss{exchange} term for identical particles $\gint$ (bottom) as defined through Eqs.~\eqref{eq:GauntQuadrupole} and~\eqref{eq:gaunt_id} in the range $10^{-3}\le|\nu_i|\le 10^4$ in increments of 0.1~dex and $0.0025 \le x \le 0.9975$ on a linear grid in~$x$ in increments of 0.0025. The values are computed using Eq.~(34) in \cite{Pradler:2020znn}. The dots indicate that we have set $\gint=0$ for $|\nu_i|\geq 10$ since the relative importance of $\gint$ with respect to $\gnonid$ is smaller than the provided precision of the table; the notation follows Tab.~\ref{tab:1}.
\label{tab:6}}
\end{table*}

\bibliography{References.bib}
\end{document}